\documentclass[a4paper,fleqn]{cas-sc}

\usepackage[authoryear,longnamesfirst]{natbib}
\usepackage{amsmath,amssymb}
\usepackage{booktabs}
\usepackage{graphicx}
\usepackage{subcaption}
\usepackage{multirow}
\usepackage{enumitem}
\usepackage{pgfplots}
\pgfplotsset{compat=1.18}
\usepackage{pgfplotstable}
\usepackage{tikz}
\usepgfplotslibrary{groupplots}
\usepackage[ruled, linesnumbered]{algorithm2e}
\SetKwInput{KwIn}{Input}
\SetKwInput{KwOut}{Output}
\SetAlgoLined
\DontPrintSemicolon
\usepackage{seqsplit}
\usepackage{listings}
\usepackage{xcolor}
\definecolor{revisionblue}{RGB}{0,0,0}
\usepackage[most]{tcolorbox}
\usepackage{colortbl}
\usepackage{float}
\usepackage{needspace}
\usepackage[section]{placeins}
\usepackage{url}

\floatstyle{plain}
\newfloat{listing}{htbp}{lop}
\floatname{listing}{Listing}

\newcommand{\ModelName}{DynaFix}
\RenewDocumentCommand \printorcid { } { }
\definecolor{diffadd}{RGB}{0,128,0}
\definecolor{diffremove}{RGB}{200,0,0}
\definecolor{ling}{RGB}{128,0,128}

\tcbset{
  answerbox/.style={
    colback=white,
    colframe=black,
    sharp corners,
    boxrule=0.8pt,
    left=5pt,right=5pt,top=5pt,bottom=5pt
  }
}

\lstdefinelanguage{diff}{
    morecomment=[f][\color{black}]{@@},
    morecomment=[f][\color{diffremove}]{-},
    morecomment=[f][\color{diffadd}]{+},
}

\newenvironment{diffpatch}{%
    \begin{tcolorbox}[colback=white, colframe=black, boxrule=0.4pt, sharp corners, left=1pt, right=1pt, top=1pt, bottom=1pt]
}{%
    \end{tcolorbox}
}

\newsavebox{\complexitytablebox}
\ExplSyntaxOn
\cs_set:Npn \__first_footerline: { }
\ExplSyntaxOff

\begin{document}
\makeatletter
\@ifundefined{@abspage@last}{}{\gdef\lastpage{\@abspage@last}}
\makeatother
\let\WriteBookmarks\relax
\def\floatpagepagefraction{1}
\def\textpagefraction{.001}

\shorttitle{DynaFix: Iterative Automated Program Repair}
\shortauthors{Z. Huang et al.}

\title[mode=title]{DynaFix: Iterative Automated Program Repair Driven by Execution-Level Dynamic Information}

\author[1]{Zhili Huang}
\ead{huangzhili@stu.cqu.edu.cn}

\author[1]{Ling Xu}
\cormark[1]
\ead{xuling@cqu.edu.cn}

\author[1]{Chao Liu}
\ead{liu.chao@cqu.edu.cn}

\author[1]{Weifeng Sun}
\ead{weifeng.sun@cqu.edu.cn}

\author[1]{Xu Zhang}
\ead{zhangx@stu.cqu.edu.cn}

\author[1]{Yan Lei}
\ead{yanlei@cqu.edu.cn}

\author[1]{Hongyu Zhang}
\ead{hyzhang@cqu.edu.cn}

\affiliation[1]{organization={School of Big Data and Software Engineering, Chongqing University},
            city={Chongqing},
            postcode={400030},
            country={China}}

\cortext[1]{Corresponding author.}

\begin{abstract}
Automated Program Repair (APR) seeks to generate correct patches for buggy programs, but recent Large Language Model (LLM)-based repair systems remain sensitive to superficial code changes and often rely on coarse test outcomes during iterative repair. We propose \emph{\ModelName{}}, an LLM-based APR framework that integrates execution-level dynamic information into a closed-loop repair workflow. {The distinguishing feature of \ModelName{} lies not in using execution-level information per se, but in repeatedly updating and exploiting such information throughout the iterative repair process.} In each repair round, \ModelName{} uses \emph{ByteTrace} to collect variable states, branch paths, and call stacks, incorporates these runtime signals into structured prompts, validates generated patches through re-execution, {and feeds the resulting updated traces into subsequent repair iterations}. This process grounds patch generation in concrete execution behavior rather than static code alone. We evaluate \ModelName{} on Defects4J v2.0, the Real-World Benchmark, and a code perturbation benchmark, Defects4J-Trans. \ModelName{} repairs 236 single-function Defects4J bugs with GPT-4 and 186 with GPT-4o, including 23 bugs not fixed by existing baselines. {On Defects4J-Trans, \ModelName{} achieves 15.6\%--32.1\% higher aggregate repair counts than ReinFix and shows higher repair retention under semantics-preserving transformations.} It also reduces token consumption by 60\%--70\% compared with state-of-the-art LLM-based APR methods. These results show that {repeatedly updating and exploiting execution-level information throughout iterative repair} can substantially improve the effectiveness, robustness, and cost efficiency of LLM-based program repair.

\end{abstract}


\maketitle
% Override the class-generated author metadata with plain author names.
\hypersetup{pdfauthor={Zhili Huang, Ling Xu, Chao Liu, Weifeng Sun, Xu Zhang, Yan Lei, Hongyu Zhang}}

\section{Introduction}
Software systems underpin nearly all critical modern domains, including manufacturing, aerospace, energy, and healthcare. As these systems grow in scale and complexity, software bugs inevitably accumulate, posing serious threats to system reliability, security, and maintainability. Program repair therefore plays a crucial role throughout the software lifecycle. Prior studies indicate that developers spend approximately 35\%–50\% of their time on manual bug diagnosis and repair~\citep{o2017debugging}. To alleviate this substantial overhead, researchers have extensively explored Automated Program Repair (APR) techniques, which aim to autonomously generate patches to correct faulty code~\citep{gazzola2018automatic}.

Early APR approaches can be broadly categorized into heuristic-based~\citep{le2011genprog, le2016history, long2016automatic, xin2017leveraging, xiong2017precise, wen2018context, jiang2018shaping}, constraint-based~\citep{xuan2016nopol, le2017s3, long2015staged, mechtaev2016angelix}, and template-based methods~\citep{martinez2016astor, long2017automatic, hua2018towards, liu2019avatar, ghanbari2019practical, jiang2019inferring}.
Heuristic-based approaches explore large patch spaces but often suffer from scalability and efficiency issues as the search space expands. Constraint-based methods offer formal guarantees but typically struggle to scale to complex programs. Template-based methods, a prominent class of traditional APR methods~\citep{benton2020effectiveness, ghanbari2019practical, liu2019tbar}, rely primarily on manually curated or mined templates. While effective in specific scenarios, their applicability is constrained by template coverage and construction cost.
More recently, deep learning–based APR methods have gained prominence by framing patch generation as a neural machine translation (NMT) task~\citep{sutskever2014sequence}, trained on large corpora of bug–fix pairs~\citep{zhang2023survey}. Despite their promising results, these methods remain highly sensitive to training data quality and distribution, and their repair effectiveness drops substantially when encountering previously unseen bug patterns or code structures~\citep{fu2022vulrepair, xia2022less, feng2024prompting}.

The rapid advancement of large language models (LLMs) has further reshaped the landscape of APR~\citep{xia2024automated, kang2025explainable, ye2024iter}. Many recent approaches adopt an iterative repair paradigm, where LLMs generate candidate patches and refine them based on feedback from prior repair attempts. Beyond purely static inputs, several studies have explored incorporating dynamic execution information, such as runtime traces and behavioral signals, to improve fault localization and patch generation~\citep{ye2022selfapr, chen2023teaching, bouzenia2023tracefixer, ding2024traced}. While these approaches achieve state-of-the-art (SOTA) performance, they still suffer from two fundamental limitations.

\begin{enumerate}[label={\arabic*)}, itemsep=3pt]

    \item \textbf{Syntactic Sensitivity of LLMs under Static Code Representations.}
    Recent empirical studies have revealed that LLMs exhibit severe sensitivity to syntactic variations in code repair tasks. \citet{le2025towards} show that variable renaming following natural programming practices can significantly degrade repair performance, suggesting an over-reliance on surface-level code patterns. Similarly, \citet{li2025evaluating} systematically evaluated 11 mainstream LLMs and found that semantically equivalent code perturbations lead to an average repair rate drop of nearly 50\%. Despite this vulnerability, many existing LLM-based APR approaches predominantly rely on static code representations, including source code snippets, exception messages, and test failure logs. Such representations fail to capture execution semantics and tend to amplify the impact of superficial syntactic changes. As a result, LLMs are encouraged to learn shallow correlations rather than underlying behavioral logic, severely limiting robustness and generalization in real-world repair scenarios.

    \item \textbf{{Limited Use of Updated Dynamic Information in Iterative Repair.}}
    SOTA LLM-based APR approaches typically employ an iterative repair strategy driven by feedback from test execution or exception reports~\citep{xia2024automated, ye2024iter, bouzenia2025repairagent, yin2024thinkrepair}. However, this feedback is largely coarse-grained, often restricted to pass/fail signals or summarized error messages. While such signals indicate final outcomes, they provide little insight into intermediate execution states or the precise causes of failure. Although recent work has explored leveraging fine-grained dynamic information such as variable states and control-flow paths~\citep{haque2025towards, ye2022selfapr, bouzenia2023tracefixer, ding2024traced}, most approaches either confine its usage to the training phase or inject it only once into the initial prompt. {Consequently, existing approaches still provide limited support for repeatedly updating and exploiting execution-level dynamic information throughout iterative repair. In particular, how to re-execute failed but still refinable candidate patches, collect execution traces that reflect their latest runtime behavior, and use the updated information to guide the next repair iteration remains insufficiently explored.}

\end{enumerate}

{To address these limitations, we propose \textbf{\ModelName{}}, an LLM-based automated program repair framework that continuously updates and exploits execution-level dynamic information within a closed-loop iterative repair process.} First, \ModelName{} collects fine-grained execution-level dynamic information, including variable states, control-flow paths, and call stacks, providing essential behavioral evidence for understanding program failures and identifying their root causes. Second, \ModelName{} constructs a progressive iterative workflow that mirrors human debugging logic. Traditional iterative methods typically expect to generate a correct patch in a single shot during each round; in contrast, when a candidate patch of \ModelName{} fails validation, the system automatically collects execution traces based on the latest code modifications. Using this updated dynamic feedback as a foundation, it guides the LLM to continue reasoning and refine the repair based on its previous attempts.

To systematically collect execution-level dynamic information, we develop \textbf{ByteTrace}, a lightweight and non-intrusive instrumentation tool that records precise, context-rich execution traces at runtime. In each repair iteration, \ModelName{} executes the buggy program, collects execution traces via ByteTrace, and augments the repair prompt with the newly observed dynamic information. The LLM then generates candidate patches conditioned on the current execution behavior. If a patch fails validation, \ModelName{} re-executes the updated program, collects revised execution traces, and refines the repair in subsequent iterations. This closed-loop process continuously grounds the LLM’s reasoning in concrete runtime evidence, closely mirroring how human developers debug programs through iterative hypothesis testing and execution tracing. By progressively incorporating fine-grained dynamic information, \ModelName{} avoids blind trial-and-error, improves repair efficiency, and enhances robustness to syntactic perturbations.

We conducted an extensive evaluation of \ModelName{} on the widely used Defects4J benchmark~\citep{just2014defects4j}, a real-world benchmark (RWB)~\citep{yin2024thinkrepair} and a code perturbation benchmark (Defects4J-Trans)~\citep{li2025evaluating}. The results demonstrate that \ModelName{} achieves a strong balance among repair effectiveness, robustness, and cost efficiency. On Defects4J v2.0, \ModelName{} achieves its peak performance with GPT-4, successfully repairing 236 single-function bugs. When using the more cost-efficient GPT-4o model, \ModelName{} repairs 186 bugs, outperforming all baselines except ReinFix and fixing 23 bugs that no other baseline methods can repair. In terms of generalization, we evaluate on the RWB. {\ModelName{} achieves numerically higher repair counts than the compared baselines across the evaluated backbone configurations, indicating promising generalization to unseen bugs.} In terms of robustness, we test on the Defects4J-Trans benchmark to measure the performance stability under syntactic variations. {The results show that \ModelName{} achieves 15.6\%--32.1\% higher aggregate repair counts than ReinFix and shows higher repair retention under semantics-preserving transformations.} In addition, \ModelName{} maintains high cost efficiency, with an average token consumption of approximately 80,058 tokens, reducing inference cost by 60\%--70\% compared to existing SOTA approaches.

Our contributions are summarized as follows:
\begin{itemize}[itemsep=3pt]
  \item {We propose \ModelName{}, an LLM-based automated program repair framework that continuously updates and exploits execution-level dynamic information within a closed-loop iterative repair process. By continuously grounding repair decisions in fine-grained runtime behavior updated across iterations, \ModelName{} achieves substantially improved robustness and generalization compared to approaches that rely solely on static contexts.}

  \item We develop ByteTrace, a lightweight and non-intrusive Java instrumentation tool that captures fine-grained execution traces, {enabling runtime feedback to be continuously updated throughout the iterative repair process.}

  \item We conduct extensive experiments on Defects4J v2.0 and additional benchmarks. The results demonstrate that \ModelName{} achieves superior repair effectiveness, repairs up to 236 bugs on Defects4J v2.0, exhibits strong robustness to semantics-preserving code perturbations, and reduces inference cost by up to 60\%--70\% compared to existing approaches. {To facilitate reproducibility, the \ModelName{} framework and all experimental datasets are publicly available at \url{https://github.com/gujiprogram/DynaFix}, while ByteTrace is released as an independent repository at \url{https://github.com/gujiprogram/ByteTrace}.}

\end{itemize}

\section{Motivation}

{This section uses the real-world Math-79 bug from Defects4J together with a controlled semantics-preserving transformation to illustrate the limitations of existing APR approaches and motivate the use of execution-level dynamic information.}

\begin{figure}[pos=htbp]
    \centering
    \includegraphics[width=\linewidth]{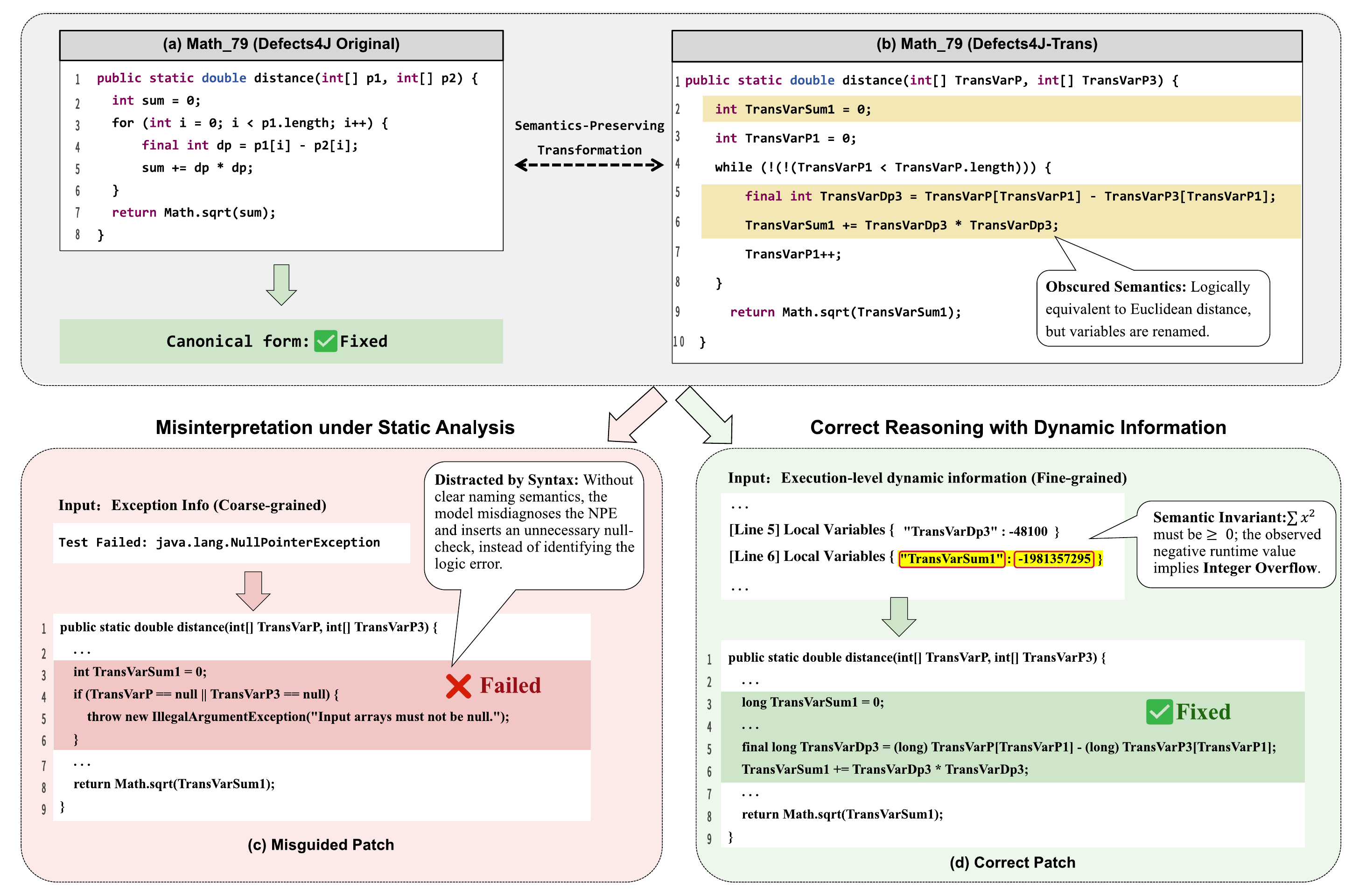}
    \caption{A motivating example using the Math-79 bug from Defects4J. (a) The original buggy code is correctly repaired under its canonical form. (b) A semantics-preserving syntactic transformation obscures the program logic. (c) When relying solely on static contexts, the model misinterprets the exception and generates an incorrect patch. (d) \ModelName{} leverages execution-level dynamic information to identify the integer overflow and produce the correct repair.}
    \label{fig:motivating_example}
\end{figure}

Figure~\ref{fig:motivating_example} illustrates the Math-79 bug, which occurs in a function that computes the Euclidean distance between two integer arrays. The failure originates from an integer overflow during the accumulation of squared differences, as shown in the statement \texttt{sum += dp * dp} {at Line 5 in Figure~\ref{fig:motivating_example}(a)}. Correctly repairing this bug requires promoting the accumulator variable from type \texttt{int} to \texttt{long}.

\begin{figure}[pos=htbp]
    \centering
    \includegraphics[width=\linewidth]{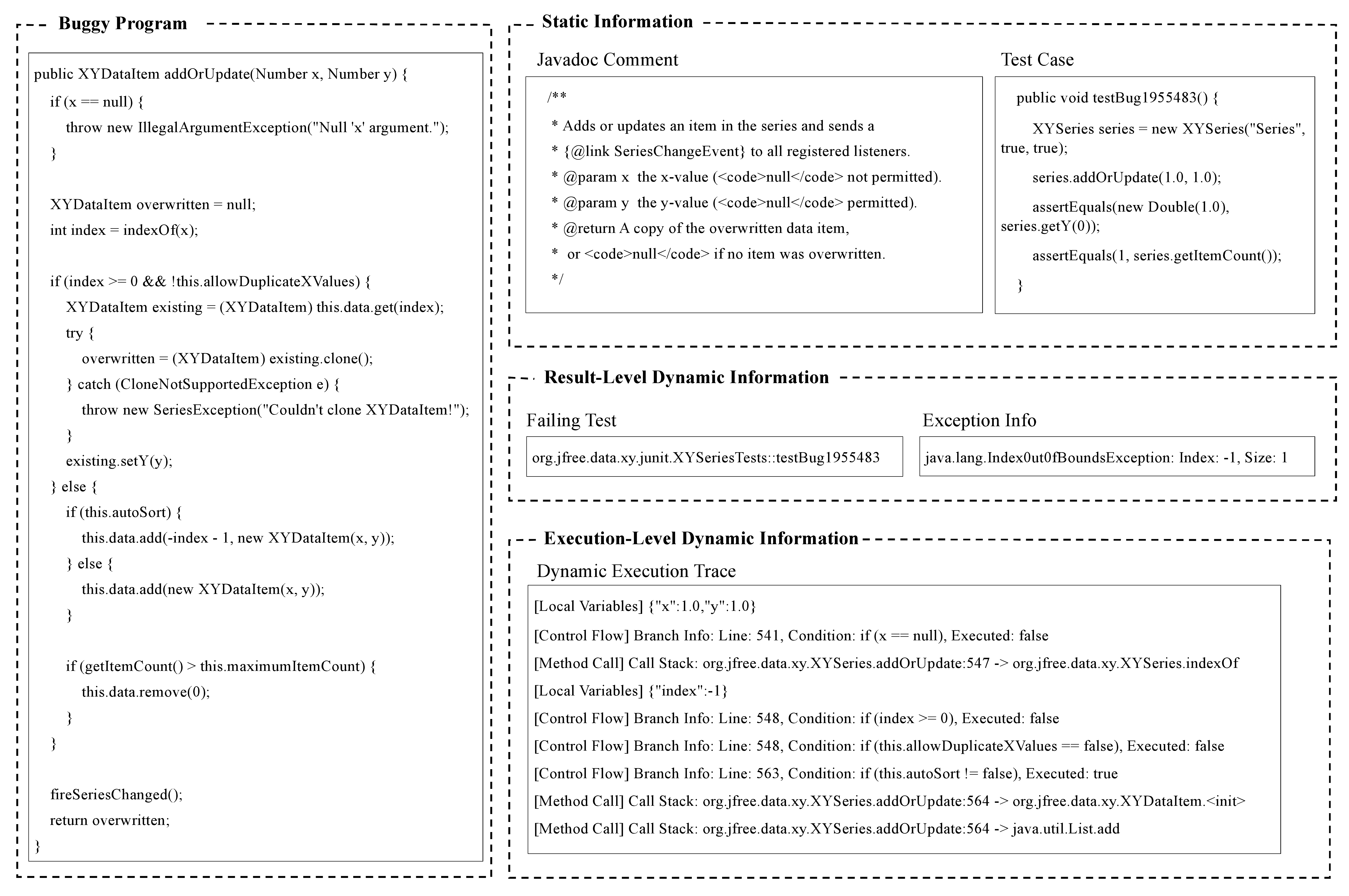}
\caption{Comparison of information types used in APR: static information, result-level dynamic information, and execution-level dynamic information.}
    \label{fig:repair_info}
\end{figure}

{Conventional APR techniques can successfully repair the original Defects4J version of this bug (Figure~\ref{fig:motivating_example}(a)). To examine sensitivity to source representation, Figure~\ref{fig:motivating_example}(b) applies a controlled semantics-preserving transformation that changes variable names and control-flow structure while preserving the underlying faulty behavior and repair target.} As shown in Figure~\ref{fig:motivating_example}(b), variable renaming and control-flow restructuring obscure semantic cues that static models often rely on{, while the corresponding overflow-prone accumulation remains at Line 6}.
In the absence of informative variable names, the model is misled by the \texttt{java.lang.NullPointerException} reported by the testing framework.
This coarse-grained exception signal fails to reveal the underlying overflow error. As a result, the model incorrectly infers that a null check is required {at Line 4}, resulting in the generation of an invalid, hallucinated patch (Figure~\ref{fig:motivating_example}(c)).

In contrast, incorporating execution-level dynamic information fundamentally changes the repair context. By examining execution traces collected at runtime, the repair process can directly observe intermediate variable states. In this example, the accumulator variable \texttt{TransVarSum1} assumes a negative value (e.g., \texttt{-1981357295}) during execution {at Line 6, as shown in Figure~\ref{fig:motivating_example}(d)}. This observation directly contradicts the mathematical property that a sum of squared values must be non-negative, providing unambiguous evidence of integer overflow. {Crucially, the negative accumulator value remains observable after this controlled source transformation, offering runtime evidence that is less affected by surface-level syntax. This evidence supports the identification of the underlying overflow and the corresponding type promotion at Lines 3 and 5 in Figure~\ref{fig:motivating_example}(d).}

The motivating example highlights a broader issue in current LLM-based APR techniques: the ineffective utilization of available information. As summarized in Figure~\ref{fig:repair_info}, the information used in APR can be categorized into three levels according to granularity.

Most existing approaches predominantly rely on static information, including faulty source code, comments, and context~\citep{wei2023copiloting, li2024enhancing, xia2023plastic, liu2019avatar}. While such information conveys program structure, it remains highly sensitive to syntactic variations and often fails to capture execution semantics. As demonstrated by the Math-79 example, static representations alone are insufficient when surface-level cues are obscured.

Other approaches incorporate dynamic signals but restrict them to result-level information, such as test pass/fail outcomes or exception stack traces. Although these signals confirm the presence of a failure, they provide limited insight into how the failure manifests during execution. In Math-79, the reported \texttt{NullPointerException} obscures the underlying overflow error, leading to incorrect fault localization due to limited visibility into the program's runtime trajectory.

These observations reveal a systematic deficiency in existing methodologies: the lack of effective and continuous utilization of execution-level information. Although some recent studies have explored leveraging variable states or control-flow information~\citep{haque2025towards, ye2022selfapr}, such information is typically used only during training or injected once into the initial prompt. Its potential to guide iterative repair remains largely unexplored. Moreover, most existing methods lack a principled feedback loop grounded in execution behavior. Static approaches typically generate patches in a single attempt, while iterative methods driven solely by result-level feedback often devolve into inefficient trial-and-error due to insufficient semantic grounding. Consequently, these methods fail to emulate the debugging process of human developers, who rely heavily on observing and reasoning about intermediate execution states.

\section{Approach}
\begin{figure}[pos=htbp]
    \centering
    \includegraphics[width=\textwidth]{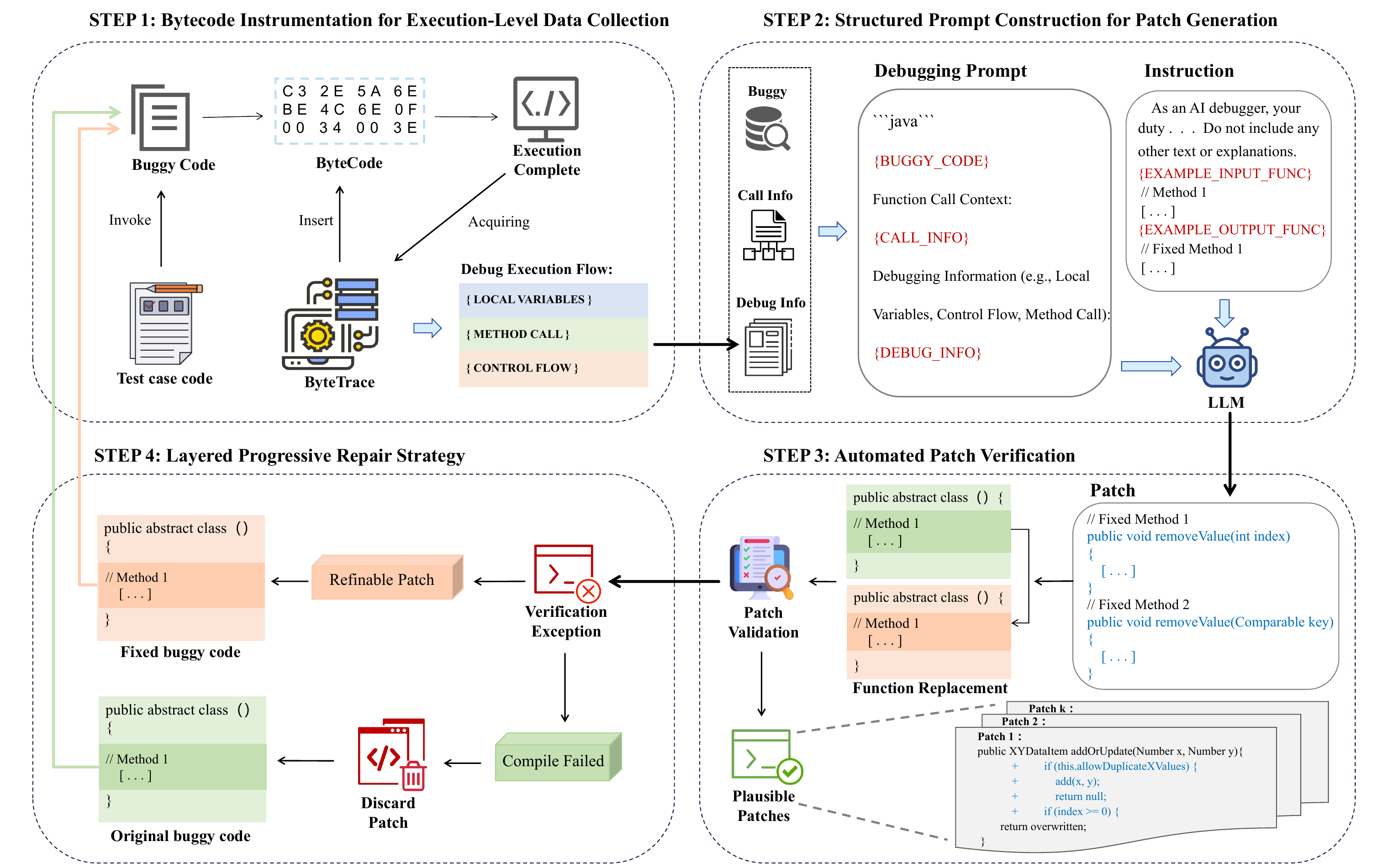}
    \caption{Architectural overview of \ModelName{}, illustrating the closed-loop process from bytecode instrumentation (Step 1) to layered progressive repair (Step 4).}
\label{fig:repair-flow}
\end{figure}

Figure~\ref{fig:repair-flow} presents an overview of \ModelName{}, an iterative APR framework driven by execution-level dynamic information.
\ModelName{} is designed to address two key challenges in LLM-based repair: (1) enriching patch generation with semantically grounded runtime evidence beyond coarse-grained exception reports, and (2) improving cost efficiency by progressively concentrating computation on promising repair trajectories.
To this end, \ModelName{} integrates non-intrusive bytecode instrumentation with a layered progressive repair (LPR) strategy, forming a closed-loop workflow in which runtime feedback is continuously incorporated to guide patch generation and refinement.

The workflow consists of four main steps:
\begin{itemize}
  \item \textbf{Step 1: Bytecode Instrumentation for Execution-Level Data Collection.}

  To mitigate the limited diagnostic capability of JUnit’s coarse-grained exception reporting, \ModelName{} performs non-intrusive bytecode instrumentation during test execution via ByteTrace. Rather than modifying source code, the framework instruments Java bytecode to capture execution-level behaviors, including variable states, control-flow paths, and call stacks, without altering program semantics. This design ensures that runtime evidence is both precise and faithful to the original execution.

  \item \textbf{Step 2: Structured Prompt Construction.}

  The collected dynamic information is integrated with the original buggy source code to construct debugging prompts. These prompts supply the LLM with concrete execution context, enabling it to reason about how and why the failure occurs, rather than relying solely on static syntax or exception messages. Instruction prompts, including explicit role definitions and one-shot examples, are incorporated to enforce output format consistency and improve patch quality.

   \item \textbf{Step 3: Automated Patch Validation.}

  Candidate patches generated by the LLM are applied to the source code and validated by running test cases. This stage provides objective feedback on patch correctness and serves as the decision point for either accepting a patch or triggering further refinement.

  \item \textbf{Step 4: Layered Progressive Repair.}
  
  Rather than repeatedly regenerating patches from scratch, DynaFix adopts an LPR strategy that distinguishes between unpromising and refinable patches at an early stage. Patches that fail to compile are immediately discarded, while those that {can be successfully compiled and executed sufficiently to obtain valid runtime feedback} are treated as refinable and iteratively improved using updated execution-level feedback. This strategy substantially reduces unnecessary LLM invocations and concentrates resources on viable repair trajectories.

\end{itemize}

\subsection{Bytecode Instrumentation for Execution-Level Data Collection}
\label{sec:bytecode-instrumentation}

A core feature of \ModelName{} is its ability to leverage execution-level dynamic information as iterative feedback to guide program repair. To collect such information from buggy programs during runtime, we integrate a non-intrusive bytecode instrumentation mechanism into the framework. This mechanism records dynamic behaviors such as local variables, method call traces, and control flow during program execution with test cases. By instrumenting bytecode instead of modifying source code directly, our approach ensures the collection of rich execution-level data without disrupting the normal execution flow of the program. Furthermore, the modular design of the instrumentation allows for scalability and automatic adaptation to benchmarks such as Defects4J.

We extended and integrated the DebugRecorder tool from DEVLoRE~\citep{feng2024integrating}, which originally utilizes JavaAgent and ASM~\citep{bruneton2002asm} to instrument Java programs and monitor local variables. In contrast, ByteTrace, our developed bytecode instrumentation tool, serves as the core data collection component of our APR framework. ByteTrace enhances the capabilities of DebugRecorder by adding support for tracing method call stacks and conditional control flow, thereby providing a more comprehensive execution context for program repair.
Concretely, ByteTrace registers a \texttt{ClassFileTransformer} during JVM startup and performs selective bytecode instrumentation at class loading time. It inserts lightweight probes at method entry and exit points, as well as at key control instructions, to capture variable states, invocation relationships, branch execution paths, and exception events. The collected runtime information is uniformly structured and tagged with thread identifiers to distinguish execution traces under concurrent settings, while preserving the original program semantics and maintaining controlled runtime overhead.
Additionally, ByteTrace modularizes the instrumentation logic, enabling efficient collection of execution-level dynamic data, which is crucial for large-scale APR experiments and benchmarking with datasets such as Defects4J~\citep{just2014defects4j}.

For information granularity selection, we adopt function-level instrumentation. Prior studies~\citep{xu2025aligning} show that function-level granularity provides sufficient context for fault localization and repair while avoiding the fragmentation problems in statement-level or variable-level tracing. This granularity also aligns well with the input--output paradigm of LLMs, which are typically trained to process and generate code at the function level. Therefore, \ModelName{} employs function-level instrumentation to balance repair effectiveness with scalability in LLM-based workflows.

\subsection{Structured Prompt Construction for Patch Generation}
\label{sec:prompt-construction}

\begin{figure}[pos=htbp]
    \centering
    \includegraphics[width=0.7\columnwidth]{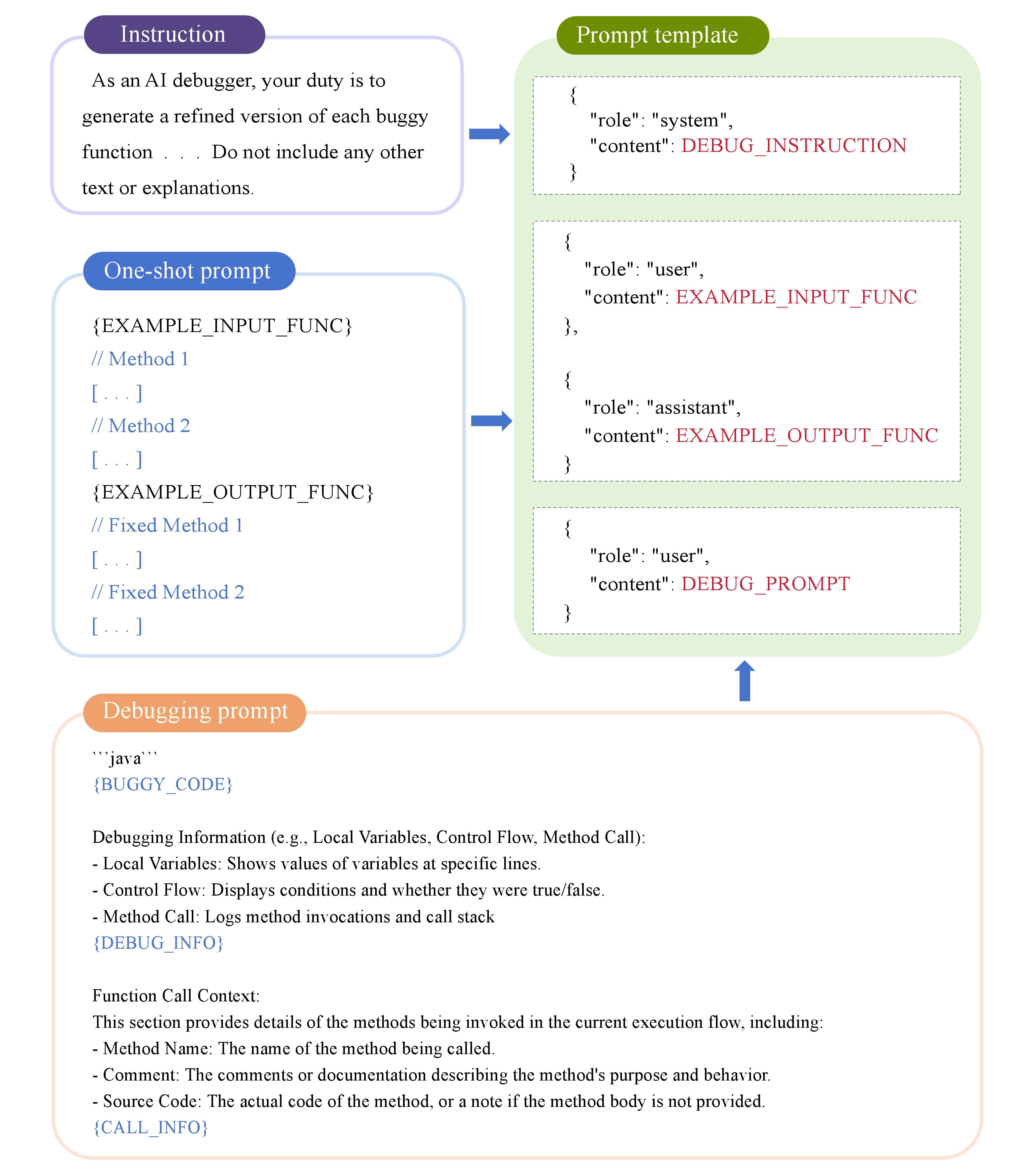}
    \caption{
    Structure of the hierarchical prompt template. A fixed input--output example is included to enforce consistent response formatting, while code details are omitted.}
\label{fig:prompt-structure}
\end{figure}

For LLM-based APR, the design of a well-structured and semantically clear prompt is critical for achieving high performance~\citep{paul2023enhancing}. Even with comprehensive bug-related information, poorly designed prompts can hinder model understanding and lead to suboptimal patches. To achieve efficient and accurate information transmission, we design a hierarchical prompt template, as illustrated in Figure~\ref{fig:prompt-structure}. This template consists of three parts: system instructions, input--output example, and debugging information.

The system instruction explicitly defines the LLM’s role as an \textit{AI debugger}, directing it to generate corrected versions of buggy functions without providing any explanatory text. This enforces role awareness and ensures the output is formatted correctly. To further guide the model, the template includes a one-shot example, inspired by recent studies~\citep{chen2024code}, which demonstrate that providing one or two examples significantly improves performance over zero-shot prompting. In this example, the input consists of a buggy function with handcrafted debugging information, while the output presents the fully repaired function. The repaired function is marked with a consistent identifier, such as \texttt{// Fixed Method X}, enabling deterministic patch extraction.

Finally, we incorporate debugging information collected by ByteTrace. This information includes: (1) the buggy function as the target for repair, (2) dynamic execution traces such as variable states, control-flow paths, and call stacks, accompanied by brief in-prompt explanations to aid model comprehension, and (3) call context, including function names, signatures, and callees, derived from dynamic call relationships within the codebase.

This structured prompt design enables the LLM to clearly identify the repair task, interpret execution-level dynamic information, and reason about call dependencies. As a result, the model is guided to generate a targeted and verifiable patch in the expected format.

\subsection{Automated Patch Validation}
\label{sec:patch-validation}
To evaluate the correctness of the generated patch, \ModelName{} replaces the faulty function in the original program with the generated fix and executes the corresponding test suite. Successful compilation and passing all tests suggest that the patch is likely correct. However, prior studies have shown that benchmark test suites often suffer from limited coverage, which can lead to the misclassification of incorrect patches as correct~\citep{tan2016anti}. To mitigate this, we follow established practices in the APR community~\citep{xia2022less, jiang2021cure} and perform manual inspection of test-passing patches. Specifically, two authors independently inspected all plausible patches {by comparing each generated patch with the corresponding developer patch and examining whether the generated modification addressed the same defect with equivalent repair semantics. Exact syntactic equivalence was not required; patches implementing an alternative but semantically equivalent repair were also considered correct. The authors first made their judgments independently. For patches on which their judgments differed, they jointly re-examined the generated patch and the corresponding developer patch and discussed the case until reaching consensus.}  {This procedure provides an additional semantic assessment beyond test-suite adequacy and reduces the risk of treating merely test-passing patches as correct.}

\subsection{Layered Progressive Repair Strategy}
\label{sec:lpr-strategy}

\begin{algorithm}[htbp]
\caption{Layered progressive repair strategy in LPR. The figure illustrates the step-by-step repair process from initial localization to progressive patch refinement.}
\label{alg:lpr}
\KwIn{
  $\mathcal{B}$: set of buggy programs with diagnostic info\\
  $B$: maximum breadth attempts per bug\\
  {$D$: maximum candidate patches per breadth, including the initial candidate}\\
  LLM: the large language model interface\\
  ByteTrace: runtime information extractor
}
\KwOut{
  Patched programs or failure reports
}
\ForEach{$\text{bug} \in \mathcal{B}$}{
  Initialize breadth attempt $b \leftarrow 0$\;
  \While{$b < B$}{
    Generate initial patch ${p_1}$ using LLM\;
    {Validate $p_1$ and obtain validation result $v_1$\;}
    \If{${p_1}$ fails to compile or {terminates with a fatal execution error}}{
      Discard ${p_1}$\;
      $b \leftarrow b + 1$\;
      continue to the next breadth attempt\;
    }
    \If{${p_1}$ passes all tests}{
      Record ${p_1}$ as successful patch and \Return\;
    }
    Initialize patch history $\mathcal{H} \leftarrow \{{(p_1, v_1)}\}$\;
    \For{{$d \leftarrow 2$ \KwTo $D$}}{
        Extract runtime info from $p_{d-1}$ via ByteTrace\;
        Generate $p_d$ via LLM using $\mathcal{H}$ and runtime info\;
        {Validate $p_d$ and obtain validation result $v_d$\;}
      \If{$p_d$ fails to compile or {terminates with} a fatal {execution} error}{
        \textbf{break} depth loop\;
      }
      \If{$p_d$ passes all tests}{
        Record $p_d$ as successful patch and \Return\;
      }
      Update patch history $\mathcal{H} \leftarrow \mathcal{H} \cup \{{(}p_d{, v_d)}\}$\;
    }
    $b \leftarrow b + 1$\;
  }
  Mark current bug as unfixable after {at most} $B \times D$ attempts\;
}
\end{algorithm}

Given the high computational cost and token-based billing model of mainstream LLM services, large-scale and multi-round invocations of models during APR can result in significant resource consumption and economic overhead. To address this issue, we introduce the LPR strategy, which incrementally explores the patch space in breadth and depth.
During the breadth phase, candidate patches are rapidly evaluated, with invalid patches discarded to save resources. {A patch is considered eligible for refinement if it can be successfully compiled and executed sufficiently to obtain valid runtime feedback. Such patches enter the depth phase, where execution-level dynamic information guides step-by-step refinement. This designation does not imply that the patch is semantically closer to the correct patch; it only indicates that the patch provides an intermediate state for bounded refinement. The breadth phase generates a wide variety of candidate patches, while the depth phase refines these eligible candidates, incorporating dynamic feedback at each step. This approach minimizes unnecessary LLM invocations and concentrates resources on refinable repair trajectories, balancing repair effectiveness with computational efficiency.} The efficiency of this strategy is discussed in Section~\ref{sec:RQ3}.

As described in Algorithm~\ref{alg:lpr}, LPR takes a set of buggy programs $\mathcal{B}$ and corresponding diagnostic information as input, and outputs either a successfully repaired program or a failure report. For each bug, the system first performs breadth-based layered repair, generating an initial patch ${p_1}$ using the LLM. If ${p_1}$ fails to compile or {terminates with a fatal execution error}, it is discarded to avoid wasting resources on invalid repair directions. If ${p_1}$ passes all test cases, the bug is considered successfully repaired, and the process terminates. {The patch history $\mathcal{H}$ stores previous candidate patches within the current refinement trajectory together with their validation results. In each depth iteration, the LLM is prompted with this history, the latest candidate patch, and the newly collected execution-level information from that candidate. The patch history helps the LLM avoid repeating previously unsuccessful modifications and encourages exploration of alternative repair directions, while execution traces from earlier iterations are not accumulated.}

{If the patches generated during breadth-based exploration do not fully repair the program but can be successfully compiled and executed sufficiently to obtain valid runtime feedback, they are considered refinable and enter the depth phase.} Here, ByteTrace collects updated execution-level information from the modified code, guiding the generation of the next patch. The patch is validated after each iteration, and if it passes all tests, the repair is deemed successful. If a patch fails compilation or {terminates with a} fatal {execution error}, the system returns to breadth-based exploration.

{LPR explores up to $B$ breadth attempts per bug, each generating at most $D$ candidate patches, including the initial candidate. If no successful patch is found after at most $B \times D$ attempts, the bug is marked as unfixable.}

\section{Experimental Setup}

\subsection{Research Questions}
This study investigates the effectiveness and characteristics of \ModelName{} through four research questions (RQs), covering repair effectiveness, robustness, efficiency, and internal mechanisms.

\begin{itemize}
    \item \textbf{RQ1: How effective is \ModelName{} under perfect and imperfect fault localization settings?}
    We evaluate \ModelName{} on the Defects4J benchmark and compare its repair performance with a set of SOTA APR techniques. In addition to the standard setting with perfect fault localization, we also assess performance under imperfect fault localization to reflect realistic debugging scenarios and examine practical applicability.

    \item \textbf{RQ2: How robust and generalizable is \ModelName{} under data leakage risks and code perturbations?}
    To evaluate generalization beyond potential training data overlap, we test \ModelName{} on the RWB dataset, which consists of bugs unseen during LLM pretraining. We further assess robustness to syntax-level code perturbations using the Defects4J-Trans benchmark and compare against state-of-the-art static retrieval-based methods.

    \item \textbf{RQ3: How do different search configurations affect the trade-off between performance and cost, and how efficient is \ModelName{} in practice?}
    We analyze the impact of different LPR search parameters on repair outcomes and LLM resource usage, and measure the runtime efficiency and overhead of \ModelName{} in practice.

    \item \textbf{RQ4: How do different feedback mechanisms and components contribute to repair effectiveness in \ModelName{}?}
    We conduct fine-grained ablation studies to examine the contribution of different types of execution-level dynamic information collected by ByteTrace and analyze how various feedback mechanisms influence LLM-based repair.
\end{itemize}

\subsection{Benchmarks}

\noindent
\textbf{Defects4J.}
Our primary evaluation is conducted on the widely used Defects4J benchmark~\citep{just2014defects4j}. Version 2.0 contains 835 real-world bugs from 17 open-source Java projects, of which five have been removed in later updates. Following prior work, we evaluate on the remaining 830 bugs. As in previous studies~\citep{xia2023automated, xu2025aligning, xia2024automated}, these bugs are categorized into 483 single-function bugs and 347 multi-function bugs. The single-function subset is further divided into v1.2 (255 bugs) and v2.0 (228 bugs). Defects4J provides buggy and fixed program versions together with corresponding test suites, enabling reproducible evaluation and fair comparison with existing APR techniques.

\noindent
\textbf{Real-World Benchmark (RWB).} To mitigate potential data leakage and evaluate generalization, we use the RWB dataset constructed by \citet{yin2024thinkrepair}. This dataset consists of bug-fixing commits created after the cutoff dates of major LLM pretraining corpora. RWB V1.0 contains 44 single-function bugs collected after October 2021, while RWB V2.0 includes 29 single-function bugs collected after March 2023. This dataset enables evaluation of repair performance on previously unseen real-world bugs.

\noindent
\textbf{Defects4J-Trans.} To evaluate robustness against syntax-level perturbations, we use the Defects4J-Trans dataset proposed by \citet{li2025evaluating}. This dataset applies semantics-preserving syntactic transformations to 483 single-function bugs from Defects4J. Due to version compatibility constraints, we successfully reproduced 467 transformed bugs. {Since the transformations preserve the underlying program behavior while altering its source-level representation, we use Defects4J-Trans as a controlled representation-robustness stress test to evaluate the sensitivity of repair capability to source representation.}

\subsection{Baselines}

\begin{table}[htbp]
    \centering
    \caption{{Baseline APR approaches and experimental configurations.}}
    \label{tab:baselines}
    {\small
    \setlength{\tabcolsep}{4pt}
    \renewcommand{\arraystretch}{1.18}
    \resizebox{\columnwidth}{!}{%
    \begin{tabular}{@{}ll>{\raggedright\arraybackslash}p{4.5cm}cc>{\raggedright\arraybackslash}p{2.8cm}>{\raggedright\arraybackslash}p{3.1cm}@{}}
        \toprule
        \textbf{Approach} & \textbf{Year/Venue} & {\textbf{Model}} & {\textbf{Temp.}} & {\textbf{Max Patch Budget}} & {\textbf{FL Input}} & {\textbf{Patch Validation}} \\
        \midrule
        \textbf{ReinFix} & ICSE'26 & {\texttt{gpt-4o-2024-05-13}} & {1.0} & {45} & {Perfect FL} & {TS + manual} \\
        \textbf{RepairAgent} & ICSE'25 & {\texttt{GPT-3.5-0125}} & {NR} & {NR} & {Perfect FL} & {TS + manual} \\
        \textbf{D4C} & ICSE'25 & {\texttt{gpt-4-0613}} & {1.0} & {10} & {Perfect FL} & {TS + manual} \\
        \textbf{GIANTREPAIR} & TOSEM'25 & {\texttt{gpt-3.5-turbo-0301} + StarCoder-15.5B + CodeLlama-7B + Llama-2-13B} & {0.8} & {500} & {Perfect FL} & {ExpressAPR + manual} \\
        \textbf{ChatRepair} & ISSTA'24 & {\texttt{gpt-3.5-turbo-0301}} & {1.0} & {200} & {Perfect FL} & {TS + manual} \\
        \textbf{ThinkRepair} & ISSTA'24 & {\texttt{gpt-3.5-turbo}} & {1.0} & {125} & {Perfect FL} & {TS + manual} \\
        {\textbf{SWE-agent}} & {NeurIPS'24} & {\texttt{gpt-4-1106-preview}} & {0.0} & {N/A} & {Issue + repo (agent FL)} & {SWE-bench TS} \\
        \textbf{Tare} & ICSE'23 & {T-Graph + Recoder decoder} & {N/A} & {100/stmt.} & {GZoltar + Ochiai} & {TS + manual} \\
        \textbf{FitRepair} & ASE'23 & {CodeT5-large (770M)} & {1.0} & {20,000} & {Perfect FL} & {TS + manual} \\
        \textbf{Repilot} & FSE'23 & {CodeT5-large (770M)} & {1.0} & {5,000} & {Perfect FL} & {TS + manual} \\
        \textbf{Hanabi} & TOSEM'22 & {XGBoost-based L2S-Cond} & {N/A} & {200/location} & {Ochiai + predicate switching} & {Prioritized TS + EvoSuite/manual} \\
        \textbf{AlphaRepair} & FSE'22 & {CodeBERT} & {N/A} & {5,000} & {Perfect FL} & {UniAPR + manual} \\
        \textbf{Recoder} & FSE'21 & {Syntax-guided neural decoder} & {N/A} & {100/stmt.} & {GZoltar + Ochiai} & {TS + manual} \\
        \textbf{TBar} & ISSTA'19 & {N/A (template-based)} & {N/A} & {NR} & {Perfect FL} & {TS + manual} \\
        \textbf{SimFix} & ISSTA'18 & {N/A (heuristic-based)} & {N/A} & {NR} & {Ochiai + test purification} & {TS + manual} \\
        \bottomrule
    \end{tabular}
    }%
    }
    \par\smallskip
    \begin{minipage}{\columnwidth}
    \footnotesize\raggedright
    {Notes: NR = not reported in the original study; N/A = not applicable; FL = fault localization; TS = validation using the developer-provided test suite. ``/location'' and ``/stmt.'' denote the maximum patch-generation budget per suspicious fault location and per suspicious statement, respectively. ``Manual'' denotes manual assessment of patch correctness against the developer patch.}
    \end{minipage}
\end{table}

Table~\ref{tab:baselines} summarizes the {15} representative SOTA APR techniques selected for comparison, covering a wide range of paradigms and publication venues.

For the standard Defects4J repair task, we compare \ModelName{} against 10 leading methods across four categories:
(1) retrieval-augmented and conversational approaches, including ReinFix~\citep{zhang2025repair}, D4C~\citep{xu2025aligning}, ThinkRepair~\citep{yin2024thinkrepair}, and ChatRepair~\citep{xia2024automated}.
(2) agent-based and hybrid analysis methods, including GIANTREPAIR~\citep{li2025giantrepair} and RepairAgent~\citep{bouzenia2025repairagent}.
(3) classic LLM-based APR methods, including FitRepair~\citep{xia2023plastic}, Repilot~\citep{wei2023copiloting}, and AlphaRepair~\citep{xia2022less}; and (4) the template-based method TBar~\citep{liu2019tbar}.

To evaluate performance under imperfect fault localization, we also include methods with public results in this setting, namely Tare~\citep{zhu2023tare}, Recoder~\citep{zhu2021syntax}, Hanabi~\citep{xiong2022l2s}, and SimFix~\citep{jiang2018shaping}.

{Since the baseline approaches were originally evaluated under different experimental configurations, fully aligning their models, sampling settings, patch budgets, fault-localization inputs, and validation procedures is infeasible. We therefore report only configurations explicitly documented in the original studies and mark unavailable information as NR without inference. In RQ1, we reuse the repair results reported by the original studies and normalize them to the common benchmark subsets, while DynaFix follows commonly adopted settings whenever possible, including a temperature of 1.0, separate comparisons under perfect and imperfect fault localization, and test-suite validation followed by manual inspection. Accordingly, the RQ1 results should be interpreted as comparisons under the published configurations of the respective approaches rather than as strictly controlled comparisons under identical settings.}

{Coding agents. In addition to the APR baselines above, we conduct a controlled comparison with two representative autonomous coding agents, RepairAgent and SWE-agent~{\hypersetup{citecolor=black}\citep{yang2024swe}}. RepairAgent is designed specifically for program repair, whereas SWE-agent was originally developed for repository-level issue resolution. We adapt SWE-agent to Defects4J while preserving its original agent--environment interaction workflow as much as possible. Unlike the comparisons based on published results above, these coding-agent experiments are reproduced under a unified experimental configuration, as described in Section~4.5.}

\subsection{Metrics and Statistical Analysis}

We adopt standard evaluation metrics used in prior APR studies. A \emph{plausible patch} is defined as a patch that passes all provided unit tests, while a \emph{correct patch} is a plausible patch that is semantically equivalent to the developer's fix. Following established practice~\citep{jiang2018shaping, liu2019tbar, xia2023plastic, zhu2021syntax, jiang2021cure, zhu2023tare, xia2022less, wei2023copiloting, zhang2023gamma, xiong2022l2s}, we compare methods based on the number of bugs correctly repaired.

{We employ statistical procedures according to the characteristics of the experimental data, with statistical significance assessed at $\alpha = 0.05$. For paired comparisons of repair effectiveness on the same bugs, we use the two-sided exact McNemar test, which is appropriate for paired binary outcomes. We additionally report the absolute difference in repair rates and its 95\% confidence interval using Newcombe's method for paired proportions. These analyses are applied in RQ2 to the DynaFix--ReinFix comparisons for which bug-level outcomes are available.}

{For distributional comparisons, we use} the Kolmogorov--Smirnov (KS) test to compare empirical distributions of two independent samples. Given samples of sizes $n$ and $m$, we build their empirical cumulative distribution functions (ECDFs). For a sample $X$ of size $n$, the ECDF $F_n(x)$ is defined as

\begin{equation}
F_n(x) = \frac{1}{n} \sum_{i=1}^{n} I(X_i \leq x)
\end{equation}

where $I(\cdot)$ is the indicator function. The KS statistic $D$ is the maximum vertical distance between the two ECDFs:

\begin{equation}
D_{n,m} = \sup_x |F_{1,n}(x) - F_{2,m}(x)|.
\end{equation}

We reject the null hypothesis of identical distributions when the $p$-value is below $\alpha = 0.05$. This test is applied in RQ4 to analyze differences in dynamic information size between single-function and multi-function bugs.

{For correlation analysis, we use} Spearman’s rank correlation coefficient $\rho$ to measure monotonic relationships between variables, such as dynamic trace size and runtime overhead. This method converts the original values $(X_i, Y_i)$ into ranks $R(X_i)$ and $R(Y_i)$, and the coefficient is computed as

\begin{equation}
\rho = \frac{\sum_{i=1}^{n} (R(X_i) - \bar{R}_X)(R(Y_i) - \bar{R}_Y)}
{\sqrt{\sum_{i=1}^{n} (R(X_i) - \bar{R}_X)^2}
\sqrt{\sum_{i=1}^{n} (R(Y_i) - \bar{R}_Y)^2}}.
\end{equation}

Here, $\rho \in [-1,1]$, where positive values indicate a positive correlation. Unlike Pearson correlation, Spearman does not assume linearity and is therefore suitable for non-normal data. We use Spearman’s $\rho$ in RQ3 to analyze the relationship between trace size and instrumentation time.
\subsection{Implementation and Configuration}

\noindent
\textbf{Configuration.}
We use GPT-4 and GPT-4o as the primary underlying LLMs for the main Defects4J evaluation, accessed through the OpenAI API~\citep{openai_api}, {while additional backbone models are used in RQ2 according to the corresponding benchmark protocols and baseline configurations}.
Following prior studies~\citep{xu2025aligning, xia2024automated},
the temperature is set to 1.0 to balance diversity and stability. For each bug, up to 35 candidate patches are generated.

\noindent
\textbf{\ModelName{}}. ByteTrace is implemented in Java to collect execution-level dynamic information at runtime, while the core repair logic of \ModelName{} is implemented in Python. Based on RQ3, we configure the LPR strategy with a maximum breadth of 7 and a depth of 5. {Under this configuration, each bug can undergo at most 35 rounds of candidate patch generation and validation, and the repair process terminates early once a valid patch is found or the exploration budget is exhausted.}

\noindent
{\textbf{Coding-Agent Comparison Configuration.} For the controlled coding-agent comparison, we evaluate DynaFix, RepairAgent, and SWE-agent on 388 bugs from Defects4J v1.2, covering both single-function and multi-function defects. Defects4J v1.2 originally contains 395 bugs; seven bugs removed in later Defects4J releases are excluded for all three methods. To ensure consistency, all methods use the same bug set, perfect fault localization, failing tests and failure messages, initial source-code context, and correctness assessment procedure. GPT-4o is used as the LLM backbone for all three methods.}

\noindent
\textbf{Fault Localization.}
To ensure fair comparison and realistic evaluation,
we use two fault localization settings.

\begin{itemize}
    \item \textbf{Perfect Fault Localization.}
    We use the perfect fault localization provided by Defects4J, isolating repair capability from localization accuracy. All baseline methods~\citep{zhang2025repair, xu2025aligning, yin2024thinkrepair, xia2024automated, li2025giantrepair, bouzenia2025repairagent, xia2023plastic, wei2023copiloting, xia2022less, liu2019tbar}
    adopt the same setting for fair comparison.

    \item \textbf{Imperfect Fault Localization.}
    To simulate realistic scenarios, we use GZoltar~\citep{campos2012gzoltar} with the Ochiai metric~\citep{abreu2006evaluation}.
    We select the top-5 most suspicious statements as repair locations.
\end{itemize}

\section{Results}
\subsection{RQ1: Effectiveness under Perfect and Imperfect Fault Localization Scenarios}

\subsubsection{\textit{Performance with perfect localization}}
We evaluate \ModelName{} on Defects4J v1.2 and v2.0 under a perfect fault localization setting, where the exact buggy function is known a priori. This setting is widely adopted in APR research to assess the upper-bound repair capability of a technique without the confounding effects of localization noise. To ensure a fair comparison, we normalize the results of all baseline methods to the same set of 483 single-function bugs used in this study.

\begin{table}[htbp]
\centering
\caption{Comparison with state-of-the-art APR methods. The table reports the number of correct patches on Defects4J v1.2 and v2.0.}
\label{tab:comparison_detailed}
\resizebox{\textwidth}{!}{
\begin{tabular}{lcccccccccccccc}
\toprule
\textbf{Project} & \textbf{\#Bugs} & \textbf{\ModelName{}$_{\text{GPT-4}}$} & \textbf{\ModelName{}$_{\text{GPT-4o}}$} & \textbf{ReinFix} & \textbf{D4C} & \textbf{ThinkRepair} & \textbf{GIANTREPAIR} & \textbf{ChatRepair} & \textbf{RepairAgent} & \textbf{FitRepair} & \textbf{Repilot} & \textbf{AlphaRepair} & \textbf{TBar} \\
\midrule
Chart & 16 & 13 & 9 & 14 & 5 & 9 & 8 & 15 & 11 & 8 & 6 & 8 & 9 \\
Closure & 93 & 28 & 25 & 38 & 25 & 19 & 32 & 37 & 24 & 29 & 22 & 23 & 16 \\
Lang & 40 & 21 & 29 & 24 & 23 & 15 & 13 & 21 & 15 & 17 & 15 & 11 & 10 \\
Math & 72 & 40 & 29 & 33 & 17 & 27 & 26 & 32 & 24 & 23 & 20 & 17 & 16 \\
Time & 17 & 3 & 4 & 4 & 4 & 3 & 1 & 3 & 2 & 3 & 2 & 3 & 2 \\
Mockito & 17 & 9 & 4 & 8 & 6 & 7 & 6 & 6 & 6 & 6 & 0 & 5 & 3 \\
\midrule
\textbf{Defects4J v1.2} & \textbf{255} & 114 & 100 & 121 & 80 & 80 & 86 & 114 & 82 & 86 & 65 & 67 & 56 \\
\midrule
Closure & 12 & 3 & 1 & 3 & 2 & - & 2 & - & 2 & 0 & 0 & 0 & - \\
Cli & 23 & 15 & 12 & 10 & 12 & - & 7 & - & 8 & 6 & 6 & 5 & - \\
Codec & 11 & 7 & 4 & 7 & 3 & - & 8 & - & 7 & 5 & 5 & 5 & - \\
Collections & 1 & 0 & 0 & 0 & 0 & - & 0 & - & 1 & 1 & 1 & 0 & - \\
Compress & 33 & 17 & 17 & 18 & 11 & - & 12 & - & 9 & 2 & 3 & 1 & - \\
Csv & 11 & 7 & 6 & 9 & 6 & - & 6 & - & 5 & 2 & 2 & 1 & - \\
Gson & 9 & 5 & 4 & 7 & 2 & - & 5 & - & 3 & 1 & 1 & 2 & - \\
JacksonCore & 13 & 7 & 8 & 8 & 6 & - & 8 & - & 4 & 3 & 3 & 3 & - \\
JacksonDatabind & 51 & 29 & 6 & 30 & 23 & - & 15 & - & 10 & 10 & 8 & 8 & - \\
JacksonXml & 4 & 1 & 1 & 1 & 1 & - & 1 & - & 1 & 0 & 0 & 0 & - \\
Jsoup & 53 & 28 & 25 & 29 & 24 & - & 18 & - & 15 & 13 & 17 & 9 & - \\
JxPath & 7 & 3 & 2 & 2 & 1 & - & 1 & - & 0 & 1 & 1 & 1 & - \\
\midrule
\textbf{Defects4J v2.0} & \textbf{228} & 122 & 86 & 124 & 91 & 90 & 83 & 48 & 65 & 44 & 47 & 35 & - \\
\midrule
\textbf{Total} & \textbf{483} & 236 & 186 & 245 & 171 & 170 & 169 & 162 & 147 & 130 & 112 & 102 & - \\
\bottomrule
\multicolumn{14}{l}{\small ``-'': indicates no results reported in the original work, or cannot be directly compared since different experimental settings.} \\
\end{tabular}
}
\end{table}

Table~\ref{tab:comparison_detailed} reports the number of correctly repaired bugs across projects. Overall, \ModelName{} demonstrates strong repair effectiveness under this setting. When instantiated with GPT-4, \ModelName{} successfully repairs 236 bugs, achieving its peak performance. Using the more cost-efficient GPT-4o backbone, it repairs 186 bugs, still outperforming most SOTA APR techniques. Notably, even the GPT-4o-based configuration remains highly competitive. It outperforms several agent-based methods, including GIANTREPAIR, which aggregates multiple LLM outputs, as well as RepairAgent. In addition, \ModelName{} consistently exceeds the performance of retrieval-augmented and conversational approaches such as ThinkRepair and ChatRepair. It also surpasses D4C, despite D4C relying on the stronger GPT-4 backbone. Compared with classic LLM-based baselines, the advantage is substantial: \ModelName{}$_{\text{GPT-4o}}$ repairs 56 more bugs than FitRepair (+43.1\%), 74 more than Repilot (+66.1\%), and 84 more than AlphaRepair (+82.4\%).

We observed that ReinFix achieved a higher number of repairs on Defects4J compared with DynaFix. {We further analyze this performance gap in Section~\ref{sec:complementarity-analysis} by examining the characteristics of the bugs exclusively repaired by the two approaches.} However, strong performance on a single benchmark does not necessarily indicate robustness or generalization across data distributions. To further assess this aspect, in RQ2 we systematically evaluate the generalization ability of each approach on the unseen RWB as well as under code perturbation settings using Defects4J-Trans.

\subsubsection{\textit{Complementarity Analysis}}\label{sec:complementarity-analysis} To assess the complementarity between \ModelName{} and existing APR approaches, we analyzed the overlap of successfully repaired bugs within the 483 single-function bugs considered in this study. Since several baseline methods do not publicly release detailed repair results, we restricted our analysis to the four strongest baselines that provide complete and reproducible repair lists, thereby ensuring the reliability of the comparison.

\begin{figure}[pos=htbp]
    \centering
    \begin{minipage}{0.85\columnwidth}
        \centering
        \begin{subfigure}[b]{0.45\textwidth}
            \centering
            \includegraphics[width=\linewidth]{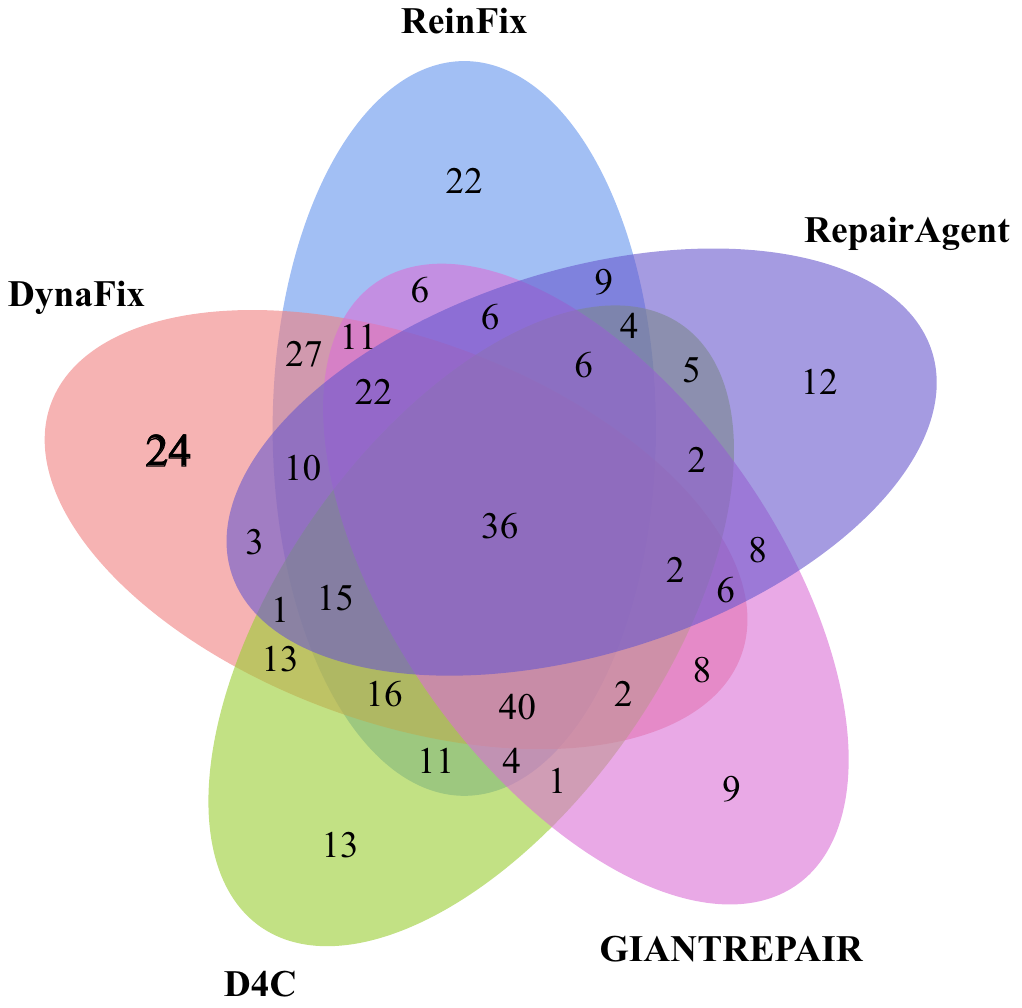}
            \caption{With top APR baselines}
            \label{fig:top}
        \end{subfigure}%
        \hfill
        \begin{subfigure}[b]{0.45\textwidth}
            \centering
            \includegraphics[width=\linewidth]{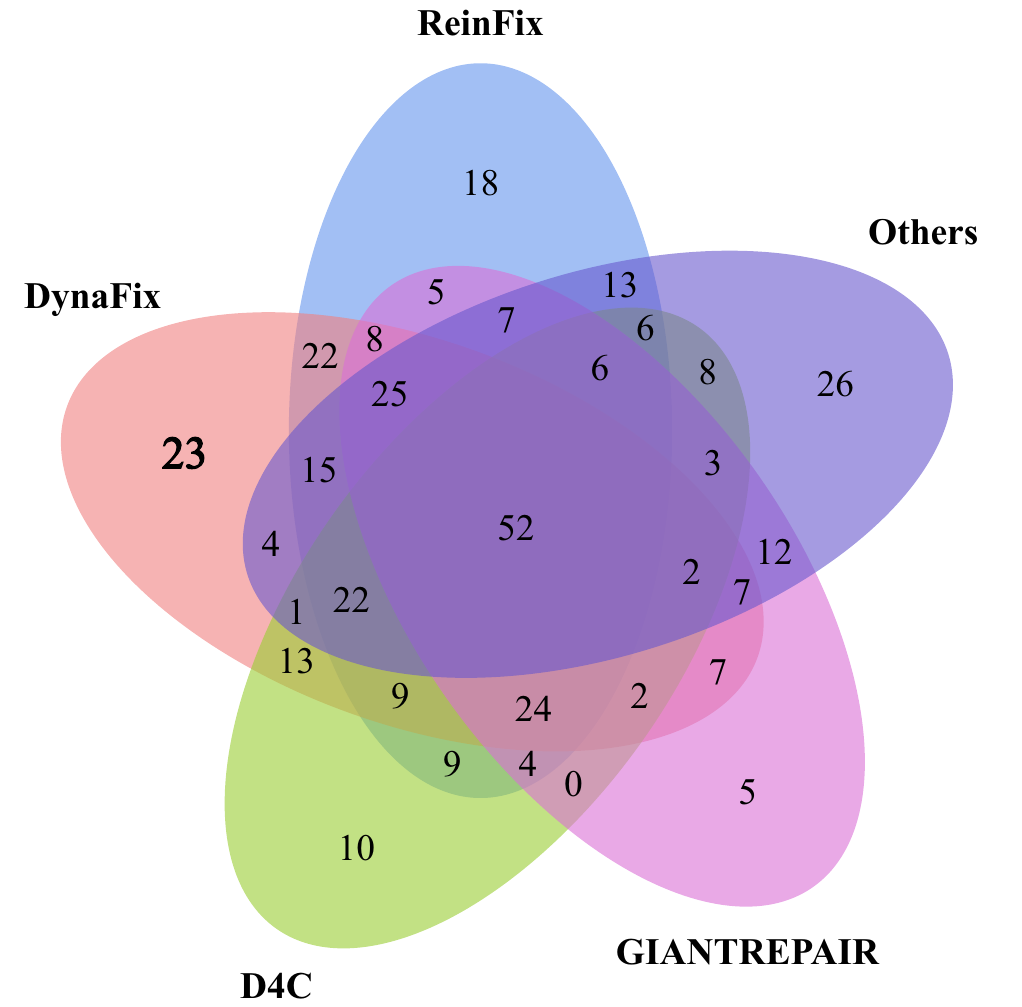}
            \caption{With all APR baselines}
            \label{fig:others}
        \end{subfigure}
    \end{minipage}
    \caption{Uniquely repaired bugs on Defects4J.}
    \label{fig:combined}
\end{figure}

\begin{listing}[tbp]
\begin{diffpatch}
\begin{lstlisting}
// Runtime context at line 192 -> Local Variable
headerBuf = [116, 101, 115, ..., 0, 0, ...]

// Developer patch
+   catch (IllegalArgumentException e) {
+      IOException ioe = new IOException("Error detected parsing the header");
+      ioe.initCause(e);
+      throw ioe;
+   }
-   currEntry = new TarArchiveEntry(headerBuf);

// DynaFix patch
+   catch (IllegalArgumentException e) {
+      throw new IOException("Invalid tar header encountered", e);
+   }
-   currEntry = new TarArchiveEntry(headerBuf);
\end{lstlisting}
\end{diffpatch}
\caption{Comparison between DynaFix patch and developer patch.}
\label{lst:unique-patch}
\end{listing}

As illustrated in Figure~\ref{fig:combined}, when compared with these four baselines, \ModelName{} uniquely repaired 24 bugs that were not addressed by any of them. Even extending the comparison to all ten baselines, \ModelName{} still contributed 23 unique repairs. These findings indicate that \ModelName{} captures a set of repair opportunities that are largely orthogonal to those addressed by existing methods. In particular, the integration of execution-level dynamic information enables \ModelName{} to resolve bugs that remain challenging for prior APR techniques, thereby providing a complementary contribution to the current APR landscape.

Listing~\ref{lst:unique-patch} shows an example bug that was repaired exclusively by \ModelName{}. The failure is triggered by the invocation of \texttt{new TarArchiveEntry(headerBuf)}. When \texttt{headerBuf} contains unparsable data, this call throws an uncaught \texttt{IllegalArgumentException}, which ultimately causes program termination.
By leveraging runtime values of local variables, \ModelName{} identifies the invalid input and accurately localizes the fault-inducing statement at line~192, enabling the generation of a correct exception-handling patch.

In contrast, existing APR techniques typically only detect the presence of an exception, without pinpointing the exact fault location or characterizing the underlying invalid input. As a result, they operate over a substantially larger search space, which reduces the likelihood of producing an effective repair.

Beyond these complementary repairs, we further analyzed why ReinFix achieves a higher overall repair count than DynaFix on the original Defects4J benchmark. Under the same GPT-4o setting, we identified 102 ReinFix-only and 43 DynaFix-only bugs. \textcolor{revisionblue}{To characterize these exclusive repairs, we manually inspected the developer patches and assigned each bug a primary defect category. We also used buggy-function LOC, cyclomatic complexity, and developer-patch size as proxies for structural and repair complexity. Table~\ref{tab:exclusive-repair-characteristics} summarizes these results.}

% [ADDED: a single table with two panels; NOT two extra figures]
\begin{table}[!h]
\centering
\begingroup
\color{revisionblue}
\arrayrulecolor{revisionblue}
\caption{\textcolor{revisionblue}{Bug characteristics and complexity of exclusive repairs on Defects4J.}}
\label{tab:exclusive-repair-characteristics}
\footnotesize
\setlength{\tabcolsep}{4pt}
\renewcommand{\arraystretch}{1.12}
\textbf{(a) Primary bug-type distribution}\par\smallskip
\begin{tabular}{@{}p{0.57\columnwidth}rr@{}}
\toprule
Primary bug type & DynaFix-only & ReinFix-only \\
\midrule
Parsing / string handling & 13 (30.2\%) & 23 (22.5\%) \\
Boundary / special-value handling & 10 (23.3\%) & 20 (19.6\%) \\
Branch / condition logic & 7 (16.3\%) & 16 (15.7\%) \\
Exception / runtime type handling & 3 (7.0\%) & 10 (9.8\%) \\
Algorithm / calculation logic & 3 (7.0\%) & 4 (3.9\%) \\
Data structure / state update & 3 (7.0\%) & 10 (9.8\%) \\
Time / environment / concurrency & 2 (4.7\%) & 6 (5.9\%) \\
API misuse / wrong method call & 1 (2.3\%) & 12 (11.8\%) \\
Other & 1 (2.3\%) & 1 (1.0\%) \\
\midrule
Total & 43 & 102 \\
\bottomrule
\end{tabular}
\par\medskip
\textbf{(b) Structural and repair complexity}\par\smallskip
\sbox{\complexitytablebox}{%
\begin{tabular}{@{}p{0.31\columnwidth}rrcc@{}}
\toprule
Metric & \shortstack{ReinFix-only\\Median [IQR]} & \shortstack{DynaFix-only\\Median [IQR]} & \shortstack{Holm-adjusted\\$p$} & \shortstack{Project-centered\\$p$} \\
\midrule
Buggy-function LOC & 25.0 [37.0] & 18.0 [16.25] & 0.3261 & 0.0302 \\
Cyclomatic complexity & 5.0 [12.0] & 5.0 [6.0] & 0.6256 & 0.1566 \\
Developer patch changed lines & 2.0 [1.0] & 3.0 [5.0] & 0.0091 & 0.3231 \\
Developer patch hunks & 1.0 [0.0] & 1.0 [0.0] & -- & -- \\
\bottomrule
\end{tabular}%
}
\usebox{\complexitytablebox}
\par\vspace{3pt}

\endgroup
\end{table}

% [ADDED: analysis paragraph 1 -- bug types]
\textcolor{revisionblue}{As shown in Table~\ref{tab:exclusive-repair-characteristics}(a), the two groups exhibit broadly similar distributions across most bug categories. For example, branch/condition logic accounts for 16.3\% and 15.7\% of DynaFix-only and ReinFix-only bugs, respectively, while boundary/special-value handling accounts for 23.3\% and 19.6\%. The most noticeable differences occur in parsing/string handling and API misuse/wrong method calls, but no dominant bug category emerges that could explain ReinFix's additional repairs. We also examined runtime-observable evidence, which was found in 40/43 DynaFix-only bugs (93.0\%) and 82/102 ReinFix-only bugs (80.4\%). However, this difference was not statistically significant (Fisher's exact test, \(p=0.080\)). Overall, no single bug-type factor sufficiently explains ReinFix's advantage.}

% [ADDED: analysis paragraph 2 -- complexity]
\textcolor{revisionblue}{Table~\ref{tab:exclusive-repair-characteristics}(b) further compares the structural and repair complexity of these exclusive repairs. The two groups show no significant difference in cyclomatic complexity. For buggy-function LOC, the pooled comparison is not significant (Holm-adjusted \(p=0.3261\)), whereas the project-centered analysis indicates a significant difference (\(p=0.0302\)). Notably, ReinFix-only bugs occur in larger functions in terms of median LOC (25 vs. 18), rather than smaller ones. Although ReinFix-only bugs have smaller developer patches in the pooled comparison (median changed lines: 2 vs. 3, Holm-adjusted \(p=0.0091\)), this difference is no longer significant after project-centering (\(p=0.3231\)). Overall, these results do not support a simple explanation that ReinFix's additional repairs are attributable to smaller or structurally simpler bugs.}

% [UNCHANGED except the former one-sentence bug-type summary was removed]
\textcolor{revisionblue}{Finally, we investigate a possible project-level explanation based on ReinFix's retrieval-augmented design. Since ReinFix retrieves repair knowledge using embedding-based cosine similarity, we examine the Top-1 retrieval similarity across Defects4J projects and its relationship with the distribution of exclusive repairs. As shown in Table~\ref{tab:retrieval-exclusive-advantage},} ReinFix's additional repairs are highly concentrated in JacksonDatabind and Closure, \textcolor{revisionblue}{with} net advantages of 24 and 15 bugs, \textcolor{revisionblue}{respectively. Together, these projects account} for 39 of ReinFix's 59 net additional repairs over DynaFix (66.1\%). Interestingly, JacksonDatabind and Closure also exhibit the highest and second-highest median retrieval \textcolor{revisionblue}{similarities} among all Defects4J projects, at 0.7257 and 0.6193, respectively. These \textcolor{revisionblue}{findings} suggest that ReinFix may particularly benefit from projects where highly similar repair examples are available, providing a plausible explanation for a substantial portion of its advantage on the original Defects4J benchmark.

\begin{table}[pos=htbp]
\caption{{Project-level retrieval similarity and exclusive-repair advantage of ReinFix over DynaFix on Defects4J. Projects are ordered by median retrieval similarity.}}
\label{tab:retrieval-exclusive-advantage}
\centering
\begingroup

\small
\setlength{\tabcolsep}{5pt}
\renewcommand{\arraystretch}{1.12}
\arrayrulecolor{black}
\begin{tabular}{@{}lccccc@{}}
\toprule
Project & $N$ & Median similarity & ReinFix-only & DynaFix-only & Net advantage \\
\midrule
\textbf{JacksonDatabind} & 51 & \textbf{0.7257} & \textbf{25} & 1 & $\boldsymbol{+24}$ \\
\textbf{Closure} & 105 & \textbf{0.6193} & \textbf{17} & 2 & $\boldsymbol{+15}$ \\
JacksonXml & 4 & 0.6144 & 1 & 1 & $0$ \\
JacksonCore & 13 & 0.6087 & 2 & 2 & $0$ \\
Mockito & 17 & 0.5927 & 5 & 1 & $+4$ \\
JxPath & 7 & 0.5857 & 0 & 0 & $0$ \\
Gson & 9 & 0.5854 & 4 & 1 & $+3$ \\
Lang & 40 & 0.5664 & 5 & 10 & $-5$ \\
Compress & 33 & 0.5647 & 5 & 4 & $+1$ \\
Codec & 11 & 0.5602 & 4 & 1 & $+3$ \\
Collections & 1 & 0.5498 & 0 & 0 & $0$ \\
Jsoup & 53 & 0.5471 & 8 & 4 & $+4$ \\
Cli & 23 & 0.5435 & 3 & 5 & $-2$ \\
Csv & 11 & 0.5365 & 3 & 0 & $+3$ \\
Time & 17 & 0.5246 & 3 & 3 & $0$ \\
Chart & 16 & 0.5180 & 5 & 0 & $+5$ \\
Math & 72 & 0.4790 & 12 & 8 & $+4$ \\
\bottomrule
\end{tabular}
\arrayrulecolor{black}
\endgroup
\end{table}
\FloatBarrier

\begin{table}[tbp]
  \centering
  \caption{Repair results without perfect fault localization. X/Y denotes X correct patches and Y plausible patches.}
  \label{tab:repair_results}
  {\small
  \begin{tabular}{lccccccc}
    \toprule
    \textbf{Project} & \textbf{DynaFix} & \textbf{GIANTREPAIR} & \textbf{Tare} & \textbf{Recoder} & \textbf{TBar} & \textbf{SimFix} & \textbf{Hanabi} \\
    \midrule
    Chart    & 9/11  & 7/10  & 11/14 & 8/14  & 7/10  & 4/5   & 1/3 \\
    Closure  & 16/32 & 16/33 & 12/23 & 15/31 & 6/10  & 5/5   & -/- \\
    Lang     & 13/22 & 12/19 & 12/19 & 9/15  & 4/11  & 6/9   & 1/1 \\
    Math     & 23/41 & 22/40 & 18/34 & 15/30 & 12/26 & 11/20 & 13/15 \\
    Time     & 2/6   & 1/3   & 2/3   & 2/2   & 1/2   & 1/1   & 2/2 \\
    Mockito  & 5/11  & 6/6   & 2/2   & 2/2   & 1/2   & -/-   & -/- \\
    \midrule
    Total    & \textbf{68/123} & 64/111 & 57/95 & 51/94 & 31/61 & 27/40 & 17/21 \\
    \midrule
    P(\%)    & 55.28 & 57.66 & 60.00 & 54.30 & 50.82 & 67.50 & 80.95 \\
    \bottomrule
  \end{tabular}
  }
\end{table}

\subsubsection{\textit{Performance with imperfect localization}} 
We further evaluated the application of \ModelName{} under a more realistic setting with imperfect fault localization. 
Specifically, {for \ModelName{},} we employed the automatic fault localization results generated by GZoltar using the Ochiai algorithm. 
{For the baselines, we used the published results reported under their respective original configurations.} 
Table~\ref{tab:repair_results} summarizes the repair results on Defects4J v1.2, which is the common benchmark used by all baselines. 
Following standard practice, we additionally report patch precision based on the original evaluation protocols of the respective methods.
The results demonstrate that \ModelName{} maintains strong repair effectiveness even in the presence of localization noise. In total, \ModelName{} successfully repaired 68 bugs, outperforming all baseline methods evaluated under the same setting. Notably, this performance exceeds that of several LLM-based APR methods that were evaluated under perfect fault localization, such as Repilot and AlphaRepair. These results indicate that \ModelName{} exhibits robust fault tolerance with respect to imperfect localization and remains effective in realistic debugging scenarios.

\begin{tcolorbox}[answerbox]

\noindent \textbf{Answer to RQ1:}
\textit{\ModelName{} demonstrates strong repair effectiveness under both ideal and realistic fault localization settings.
Under perfect fault localization, the GPT-4-based configuration achieves {DynaFix's highest repair count, repairing 236 bugs}, while the more cost-efficient GPT-4o variant still repairs 186 bugs, {with a repair count higher than all compared baseline results except ReinFix}.
The observed performance gap with ReinFix is further {examined through complementarity and project-level retrieval analyses}.
In addition, \ModelName{} contributes {23 uniquely repaired bugs}, highlighting its complementarity with existing approaches.
Under imperfect fault localization, \ModelName{} {repairs 68 bugs, achieving the highest repair count among the compared results and remaining effective under localization noise}.}

\end{tcolorbox}

\subsection{RQ2: Generalization and Robustness Evaluation}

In RQ1, ReinFix exhibits strong performance on the original Defects4J benchmark. However, since Defects4J has been extensively used in the training corpora of contemporary LLMs, performance on this benchmark alone may not be sufficient to rule out potential overfitting. To determine whether \ModelName{} exhibits better transfer behavior rather than relying on memorized patterns, we systematically evaluate its generalization and robustness on two complementary benchmarks: the RWB and the code perturbation benchmark Defects4J-Trans.

\begin{table}[tbp]
\centering
\caption{Repair results (correct fixes) for \ModelName{} and baselines on RWB V1.0 (44 cases) and V2.0 (29 cases).}
\label{tab:repair_results_with_line}
\resizebox{\textwidth}{!}{
\begin{tabular}{lcccccccccccc}
\toprule
\textbf{Benchmark} & \multicolumn{6}{c}{\textbf{RWB V1.0 (44 bugs)}} & \multicolumn{6}{c}{\textbf{RWB V2.0 (29 bugs)}} \\
\cmidrule(lr){2-7} \cmidrule(l){8-13}

\textbf{LLM} & \multicolumn{2}{c}{\textbf{GPT-4}} & \multicolumn{3}{c}{\textbf{GPT-3.5}} & \textbf{CodeBERT} & \multicolumn{2}{c}{\textbf{GPT-4-Turbo}} & \multicolumn{3}{c}{\textbf{DeepSeek-Coder}} & \textbf{CodeBERT} \\
\cmidrule(lr){2-3} \cmidrule(lr){4-6} \cmidrule(lr){7-7} \cmidrule(lr){8-9} \cmidrule(lr){10-12} \cmidrule(l){13-13}

\textbf{APR Tool} & \textbf{DynaFix} & ReinFix & \textbf{DynaFix} & ReinFix & ThinkRepair & AlphaRepair & \textbf{DynaFix} & ReinFix & \textbf{DynaFix} & ReinFix & ThinkRepair & AlphaRepair \\

\midrule

Cli & 4 & 4 & 4 & 4 & 4 & 3 & 4 & 4 & 4 & 4 & 4 & 3 \\
Codec & 5 & 3 & 3 & 3 & 3 & 1 & 3 & 3 & 2 & 2 & 1 & 1 \\
Collections & 1 & 1 & 1 & 1 & 1 & 0 & - & - & - & - & - & - \\
Compress & 2 & 2 & 2 & 2 & 1 & 1 & 2 & 1 & 2 & 1 & 0 & 0 \\
Csv & 1 & 1 & 1 & 1 & 1 & 0 & - & - & - & - & - & - \\
Jsoup & 9 & 7 & 8 & 6 & 6 & 2 & 4 & 4 & 3 & 2 & 2 & 1 \\
Lang & 4 & 3 & 4 & 3 & 3 & 2 & 5 & 3 & 6 & 3 & 3 & 1 \\
\midrule
\textbf{Total} & \textbf{26} & 21 & \textbf{23} & 20 & 19 & 9 & \textbf{18} & 15 & \textbf{17} & 12 & 10 & 6 \\
\bottomrule
\end{tabular}
}
\end{table}

\subsubsection{\textit{Generalization on Real-World Benchmarks}}
We first evaluate the generalization capability of \ModelName{} on the RWB, which consists of bugs collected after the training cutoffs of widely used LLMs. Following the experimental protocol of ThinkRepair, we evaluate \ModelName{} using GPT-3.5 and DeepSeek-Coder as backbone models. {Note that RWB V1.0 and RWB V2.0 collect bug cases after the training cutoff dates of GPT-3.5 and DeepSeek-Coder, respectively. Therefore, the RWB dataset can further mitigate the data leakage risks of LLMs during the repair process.} 
As shown in Table~\ref{tab:repair_results_with_line}, on RWB V1.0, the GPT-3.5-based \ModelName{} repairs 23 bugs, outperforming ReinFix (20) and ThinkRepair (19). On RWB V2.0, when powered by DeepSeek-Coder, \ModelName{} successfully repairs 17 bugs, exceeding ReinFix (12) and ThinkRepair (10). These results indicate that \ModelName{} remains effective on unseen, real-world bugs even when evaluated under strict conditions designed to control for data contamination.

To further assess performance stability across stronger foundation models and to align with the best reported configurations of ReinFix, we additionally evaluate \ModelName{} using GPT-4 and GPT-4-Turbo. Considering the temporal alignment between the release of RWB V1.0/V2.0 and the respective model training cutoffs, GPT-4 is used for RWB V1.0 and GPT-4-Turbo for RWB V2.0.
As shown in Table~\ref{tab:repair_results_with_line}, {on RWB V1.0, \ModelName{} with GPT-4 repairs 26 of 44 bugs (59.09\%), compared with 21 (47.73\%) for ReinFix, corresponding to an absolute improvement of 11.36 percentage points. Bug-level paired analysis identifies eight DynaFix-only repairs and three ReinFix-only repairs. On RWB V2.0, \ModelName{} with GPT-4-Turbo repairs 18 of 29 bugs (62.07\%), compared with 15 (51.72\%) for ReinFix, corresponding to an absolute improvement of 10.34 percentage points, with five DynaFix-only and two ReinFix-only repairs.}

{Despite the consistently higher repair rates, the two-sided exact McNemar tests do not reach statistical significance on either RWB V1.0 ($p=0.227$) or RWB V2.0 ($p=0.453$). The corresponding 95\% confidence intervals for the paired repair-rate differences are [$-3.56$, $25.36$] and [$-7.95$, $27.44$] percentage points, respectively. These results indicate a consistent numerical advantage for \ModelName{} across both matched configurations, although the current RWB sample sizes do not provide sufficient statistical evidence to establish superiority over ReinFix. Bug-level repair identities are unavailable for the other ReinFix configurations reported in Table~\ref{tab:repair_results_with_line}; therefore, those comparisons are treated descriptively and are not included in the paired significance analysis.}

\begin{table}[htbp]
  \centering
  \caption{Repair results (correct fixes) of \ModelName{} and ReinFix on the Defects4J-Trans benchmark.}
  \label{tab:d4j_trans_comparison_optimized}
  {\small
  \begin{tabular}{l c cc cc}
    \toprule
    \textbf{Project} & \textbf{\#Bugs} & \textbf{DynaFix$_{\text{DeepSeek-V3.2}}$} & \textbf{ReinFix$_{\text{DeepSeek-V3.2}}$} & \textbf{DynaFix$_{\text{GPT-4o}}$} & \textbf{ReinFix$_{\text{GPT-4o}}$} \\
    \midrule

    Chart           & 16  & 11 & \textbf{12} & 8  & \textbf{12} \\
    Cli             & 23  & \textbf{15} & 7  & \textbf{10} & 7  \\
    Closure         & 102 & \textbf{44} & 32 & \textbf{18} & 15 \\
    Codec           & 11  & \textbf{6}  & 4  & \textbf{4}  & 1  \\
    Collections     & 1   & 0  & 0  & 0  & 0  \\
    Compress        & 32  & \textbf{21} & 9  & \textbf{13} & 12 \\
    Csv             & 11  & \textbf{9}  & 7  & 4  & \textbf{5}  \\
    Gson            & 9   & 0  & \textbf{1}  & \textbf{5}  & 0  \\
    JacksonCore     & 13  & 7  & 7  & 6  & \textbf{10} \\
    JacksonDatabind & 50  & 1  & \textbf{9}  & \textbf{15} & 6  \\
    JacksonXml      & 4   & 1  & 1  & 1  & 1  \\
    Jsoup           & 50  & \textbf{30} & 28 & \textbf{20} & 17 \\
    JxPath          & 7   & 0  & \textbf{1}  & 0  & \textbf{1}  \\
    Lang            & 35  & \textbf{17} & 14 & \textbf{13} & 12 \\
    Math            & 71  & \textbf{42} & 22 & \textbf{30} & 27 \\
    Mockito         & 16  & 6  & \textbf{7}  & 6  & \textbf{8}  \\
    Time            & 16  & \textbf{4}  & 1  & \textbf{3}  & 1  \\

    \midrule
    \textbf{Total}  & \textbf{467} & \textbf{214} & 162 & \textbf{156} & 135 \\
    \bottomrule
  \end{tabular}
  }
\end{table}

\subsubsection{\textit{Robustness under Code Perturbation}}
To evaluate robustness against syntactic variation, we adopt the Defects4J-Trans benchmark, which applies semantics-preserving code transformations to the original Defects4J programs. As shown in Table~\ref{tab:d4j_trans_comparison_optimized}, {\ModelName{} achieves higher aggregate repair performance on the transformed programs.} When using DeepSeek-V3.2 as the backbone model, \ModelName{} repairs 214 bugs, outperforming ReinFix by 52 bugs (+32.1\%). Similarly, with GPT-4o, \ModelName{} repairs 156 bugs, exceeding ReinFix by 21 bugs (+15.6\%).
{However, the numbers of bugs vary considerably across projects, and ReinFix achieves higher repair counts on several individual projects. Therefore, aggregate repair counts may be disproportionately influenced by larger projects and do not fully characterize repair stability before and after code transformations.}

\begin{figure}[pos=htbp]
    \centering
    \includegraphics[width=\linewidth]{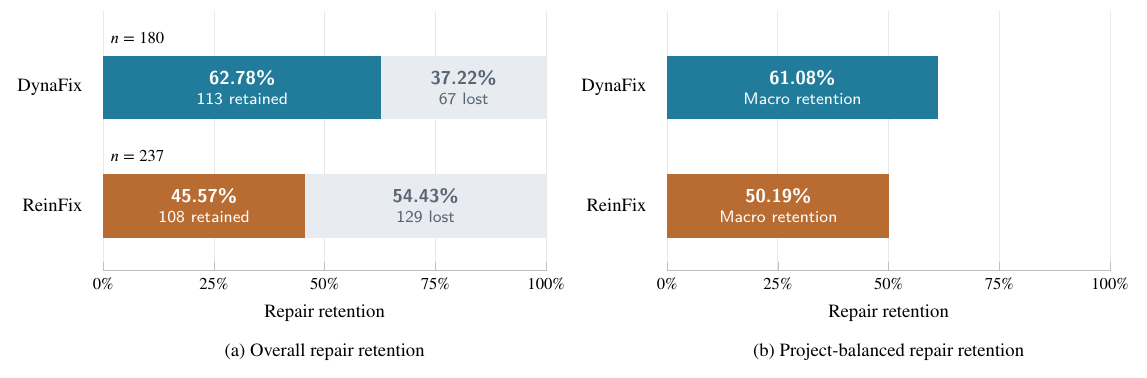}
    \caption{{Repair retention under semantics-preserving transformations. (a) Overall retention of originally successful repairs across all 467 paired bugs. (b) Project-balanced retention, computed as the equal-weight macro-average across 16 comparable projects; Collections is excluded because both methods have zero original correct repairs, making retention undefined.}}
    \label{fig:repair-retention}
\end{figure}

{To further characterize robustness under semantics-preserving transformations, we conduct a paired repair-retention analysis on the same 467 original--transformed bug pairs. Repair retention measures the proportion of bugs correctly repaired in the original programs that remain correctly repaired after transformation. As shown in Figure~\ref{fig:repair-retention}(a), \ModelName{} retains 113 of 180 originally successful repairs (62.78\%), compared with 108 of 237 (45.57\%) for ReinFix, yielding a 17.21 percentage-point higher retention rate. Correspondingly, the repair loss rates are 37.22\% and 54.43\%, respectively, indicating that \ModelName{} preserves more of its original repair capability after code transformation.}

{To reduce the influence of unequal project sizes, we further report a project-balanced repair retention, computed as the equal-weight macro-average across 16 comparable projects. As shown in Figure~\ref{fig:repair-retention}(b), \ModelName{} achieves 61.08\%, exceeding ReinFix's 50.19\% by 10.89 percentage points. This result shows that \ModelName{}'s higher repair retention persists after balancing project sizes.}

\subsubsection{\textit{Root Cause Analysis of Performance Disparity}}
To investigate the underlying cause of this performance disparity, we analyze the retrieval behavior of ReinFix under code perturbation. Specifically, we compare the similarity distributions of retrieved patch fragments for the same 467 bugs on Defects4J and Defects4J-Trans.

\begin{figure}[pos=htbp]
    \centering
    \begin{subfigure}[b]{0.479\linewidth}
        \centering
        \includegraphics[width=\linewidth]{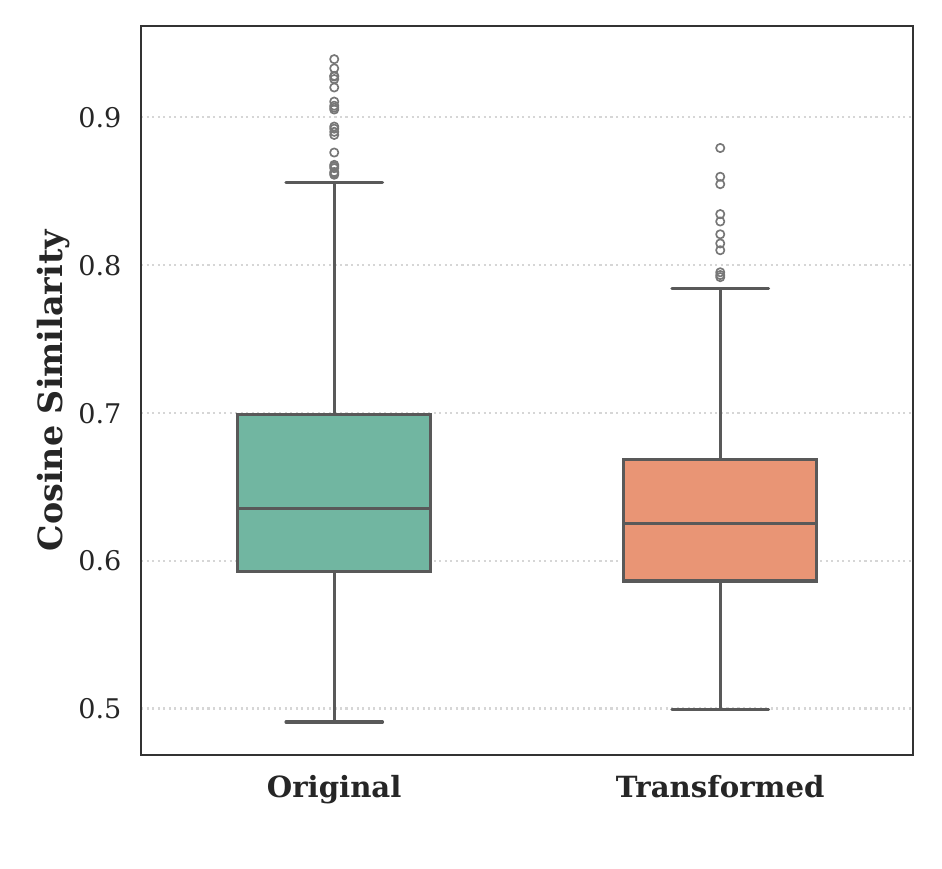}
        \caption{Distribution shift}
        \label{fig:sim_box}
    \end{subfigure}
    \hfill
    \begin{subfigure}[b]{0.48\linewidth}
        \centering
        \includegraphics[width=\linewidth]{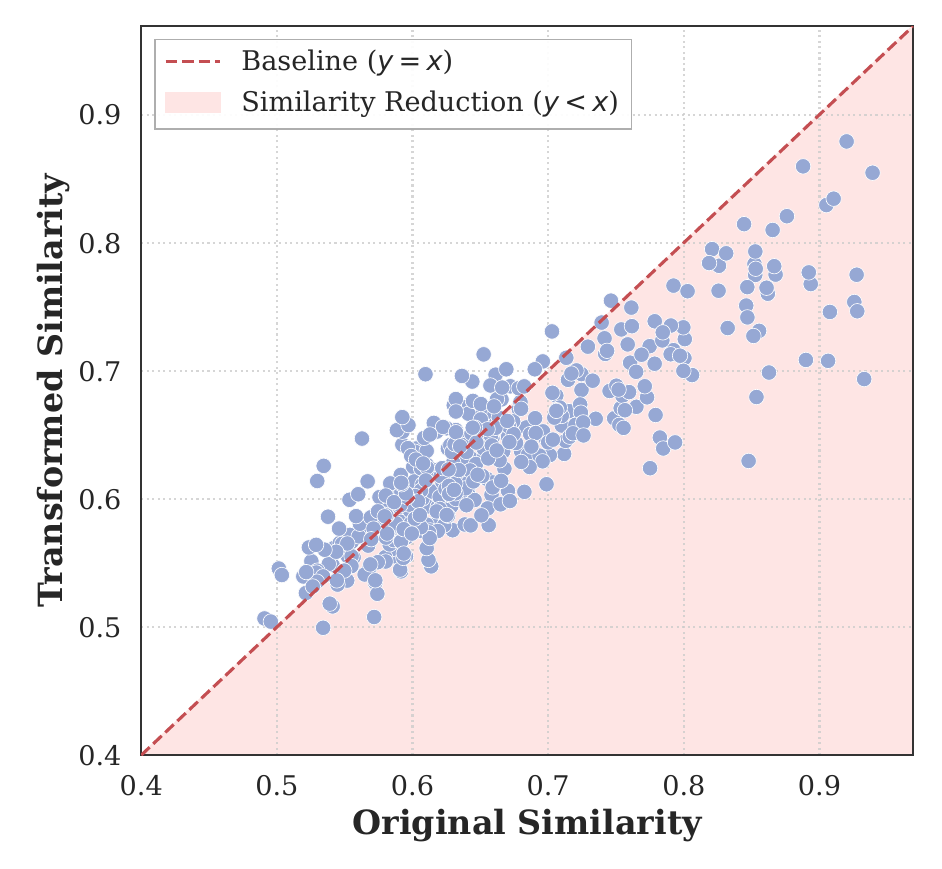}
        \caption{Paired similarity analysis}
        \label{fig:sim_scatter}
    \end{subfigure}
    \caption{Impact of code perturbation on retrieval similarity. (a) Comparison of cosine similarity distributions, showing a downward shift after transformation. (b) Paired sample analysis where points below the $y=x$ baseline (shaded area) indicate a degradation in retrieval confidence due to syntactic changes.}
    \label{fig:similarity_analysis}
\end{figure}

Figure~\ref{fig:similarity_analysis} shows a pronounced downward shift in retrieval similarity after perturbation. Statistical analysis reveals that 307 out of 467 bugs (65.74\%) experience a similarity decrease, which is confirmed to be highly significant by a Wilcoxon signed-rank test ($p = 2.33 \times 10^{-21}$). Notably, the impact is most severe for high-confidence retrievals: the number of cases with similarity above 0.75 drops from 78 to 30 after transformation, indicating that more than 60\% of high-quality retrieval contexts are lost.

These findings explain the sensitivity of static retrieval-based methods to syntactic variation. \textcolor{revisionblue}{In contrast, DynaFix leverages execution-level dynamic traces to guide the repair process, reducing its reliance on surface-level syntactic features. Although semantics-preserving transformations may alter the representation and structure of execution traces, DynaFix can still utilize behaviorally relevant information to mitigate the impact of syntactic variations and maintain robust repair performance under the evaluated code perturbations.}

\begin{tcolorbox}[answerbox]

\noindent \textbf{Answer to RQ2:}
\textit{
{\ModelName{} shows effective repair performance on post-cutoff real-world bugs and higher repair retention under semantics-preserving transformations.}
On the RWB, it {achieves numerically higher repair counts than the compared baselines across the evaluated backbone configurations}, indicating effective generalization to unseen bugs.
{On Defects4J-Trans, \ModelName{} achieves higher aggregate repair performance than ReinFix and shows higher repair retention under semantics-preserving transformations. Specifically, \ModelName{} achieves an overall retention rate of 62.78\%, compared with 45.57\% for ReinFix, while the project-balanced retention rates are 61.08\% and 50.19\%, respectively.}
Further analysis {shows that ReinFix's retrieval similarity degrades significantly under code perturbation, providing a plausible explanation for the sensitivity of retrieval-based repair to syntactic variation}.
{Overall, \ModelName{} shows higher repair retention, including higher project-balanced retention, under the evaluated semantics-preserving transformations.}
}

\end{tcolorbox}

\subsection{RQ3: Evaluation of Performance-Cost Trade-off and Runtime Efficiency}
\label{sec:RQ3}
To assess the practical deployability of \ModelName{} in realistic software maintenance settings, we examine its performance–cost trade-off and runtime efficiency from three complementary perspectives: (i) sensitivity to LPR search parameters, (ii) runtime overhead introduced by dynamic information acquisition, and (iii) token consumption and economic efficiency.

\subsubsection{\textit{Impact of LPR Search Parameters on Repair Performance}}
We first analyze how the key hyperparameters of the LPR strategy---namely the search breadth ($B$) and search depth ($D$)---affect repair effectiveness and computational cost. Since exhaustively exploring the joint parameter space is computationally prohibitive, we conduct a grid search on a representative subset of 255 single-function bugs from Defects4J v1.2.

{To quantify the monetary cost under different search configurations, we use Average Repair Cost (USD/bug). For each configuration of $B$ and $D$, we calculate the total cost of all LLM invocations over the 255 evaluated bugs, including both input and output token consumption, and divide it by the total number of bugs. Formally,}
\begingroup
\[
C_{\mathrm{avg}} = \frac{1}{N} \sum_{i=1}^{N} \left( \frac{T_i^{\mathrm{in}}}{10^6}P_{\mathrm{in}} + \frac{T_i^{\mathrm{out}}}{10^6}P_{\mathrm{out}} \right),
\]
\endgroup
{where $N=255$, $T_i^{\mathrm{in}}$ and $T_i^{\mathrm{out}}$ denote the total numbers of input and output tokens consumed for bug $i$, respectively, and $P_{\mathrm{in}}$ and $P_{\mathrm{out}}$ denote the corresponding API prices per million input and output tokens. The metric is measured in USD per bug and includes the cost incurred for all evaluated bugs regardless of whether they are ultimately repaired. A higher Average Repair Cost indicates greater LLM inference expenditure under the corresponding search configuration, whereas a lower value indicates lower monetary cost.}

\begin{figure}[pos=htbp]
    \centering
    \begin{minipage}{\linewidth}
        \centering
        \begin{subfigure}[t]{0.47\linewidth}
            \centering
            \begin{tikzpicture}
            \scriptsize
            \begin{axis}[
                legend style={font=\scriptsize},
                legend cell align={left},
                title={Depth = 3},
                width=\linewidth,
                height=0.22\textheight,
                xlabel={Search Breadth},
                ylabel={{Avg. Repair Cost (USD/bug)}},
                ymin=0, ymax=0.8,
                ybar,
                bar width=10pt,
                ymajorgrids=true,
                xtick=data,
                xtick={1,3,5,7,10},
                xticklabels={1,3,5,7,10},
                legend pos=north west,
            ]
            \addplot+[ybar, draw=gray!40!teal!40, fill=gray!40!teal!40]
            coordinates {(1,0.13)(3,0.20)(5,0.31)(7,0.38)(10,0.56)};
            \addlegendentry{{Average Repair Cost}}
            \end{axis}

            \begin{axis}[
                legend style={font=\scriptsize},
                legend cell align={left},
                width=\linewidth,
                height=0.22\textheight,
                ylabel={Plausible Patch},
                ymin=50, ymax=140,
                axis y line*=right,
                axis x line=none,
            ]
            \addplot[color=blue!90!black, mark=*, mark options={draw=none, fill=blue!90!black}]
            coordinates {(1,57)(3,95)(5,93)(7,109)(10,103)};
            \addlegendentry{Plausible Patch}
            \draw[dashed, black!50] (axis cs:7,50) -- (axis cs:7,140);
            \end{axis}
            \end{tikzpicture}
            \caption{\footnotesize Impact of Search Breadth on Fix Performance and {Average Repair Cost}}
        \end{subfigure}%
        \hspace{0.025\linewidth}
        \begin{subfigure}[t]{0.47\linewidth}
            \centering
            \begin{tikzpicture}
            \scriptsize
            \begin{axis}[
                legend style={font=\scriptsize},
                legend cell align={left},
                title={Breadth = 5},
                width=\linewidth,
                height=0.22\textheight,
                xlabel={Search Depth},
                ylabel={{Avg. Repair Cost (USD/bug)}},
                ymin=0, ymax=0.8,
                ybar,
                bar width=10pt,
                ymajorgrids=true,
                xtick=data,
                xtick={1,3,5,7,10},
                xticklabels={1,3,5,7,10},
                legend pos=north west,
            ]
            \addplot+[ybar, draw=gray!40!teal!40, fill=gray!40!teal!40]
            coordinates {(1,0.24)(3,0.32)(5,0.36)(7,0.39)(10,0.59)};
            \addlegendentry{{Average Repair Cost}}
            \end{axis}

            \begin{axis}[
                legend style={font=\scriptsize},
                legend cell align={left},
                width=\linewidth,
                height=0.22\textheight,
                ylabel={Plausible Patch},
                ymin=50, ymax=140,
                axis y line*=right,
                axis x line=none,
            ]
            \addplot[color=blue!90!black, mark=*, mark options={draw=none, fill=blue!90!black}]
            coordinates {(1,84)(3,91)(5,109)(7,111)(10,103)};
            \addlegendentry{Plausible Patch}
            \draw[dashed, black!50] (axis cs:5,50) -- (axis cs:5,140);
            \end{axis}
            \end{tikzpicture}
            \caption{\footnotesize Impact of Search Depth on Fix Performance and {Average Repair Cost}}
        \end{subfigure}
    \end{minipage}%
    \caption{Impact of Search Depth and Breadth on Plausible Patches and {Average Repair Cost (USD/bug)} in the LPR Strategy.}
    \label{fig:lpr-depth-breadth-impact}
\end{figure}

Figure~\ref{fig:lpr-depth-breadth-impact} illustrates the impact of varying $B$ and $D$ on the number of plausible patches and the corresponding {Average Repair Cost}. When the search depth is fixed at $D=3$, {increasing the search breadth generally improves the number of plausible patches, although the trend is not strictly monotonic. In particular, a small decrease is observed when $B$ increases from 3 to 5, followed by a further improvement at $B=7$.} Beyond this point, the improvement becomes marginal, while the {Average Repair Cost} continues to grow approximately linearly.
A similar pattern is observed when fixing the search breadth at $B=5$. Increasing the search depth yields moderate performance improvements up to $D=5$, after which the number of plausible patches plateaus. Further increases in depth primarily introduce redundant or lower-quality patch candidates, accompanied by rising computational cost.

These results suggest a clear diminishing-returns effect once the search space exceeds a moderate scale ($B>7$ or $D>5$). This behavior is likely constrained by the generative diversity and sampling limitations of the underlying LLM. Expanding the search space beyond this regime does not substantially improve patch recall, but instead incurs unnecessary overhead. Based on this analysis, we adopt $B=7$ and $D=5$ as the default configuration of \ModelName{}, striking a balanced trade-off between repair effectiveness and efficiency.

\subsubsection{{\textit{Analysis of LPR Refinement Behavior}}}
\label{sec:lpr-refinement-behavior}
{To further understand the refinement behavior of LPR under the default configuration, we analyze the repair trajectories from the main Defects4J experiments on the 483 single-function bugs. Specifically, we examine the depth at which correct patches are obtained, the evolution of test outcomes during successful refinement, and the termination of non-productive trajectories. Table~\ref{tab:lpr-refinement-depth} summarizes the distribution of correct repairs across LPR depth levels.}

\begin{table}[pos=htbp]
\centering
\caption{{Distribution of correct repairs across LPR depth levels.}}
\label{tab:lpr-refinement-depth}
\begingroup
\small
\setlength{\tabcolsep}{3pt}
\renewcommand{\arraystretch}{1.15}
\arrayrulecolor{black}
\makebox[\linewidth][c]{%
\begin{tabular}{@{}lrrrrrrr@{}}
\toprule
Model & $d=1$ & $d=2$ & $d=3$ & $d=4$ & $d=5$ & Refined ($d\ge2$) & Total \\
\midrule
\ModelName{}$_{\text{GPT-4}}$ & 186 (78.81\%) & 22 (9.32\%) & 20 (8.47\%) & 7 (2.97\%) & 1 (0.42\%) & \textbf{50 (21.19\%)} & 236 \\
\ModelName{}$_{\text{GPT-4o}}$ & 121 (65.05\%) & 26 (13.98\%) & 25 (13.44\%) & 12 (6.45\%) & 2 (1.08\%) & \textbf{65 (34.95\%)} & 186 \\
\bottomrule
\end{tabular}}
\arrayrulecolor{black}
\par\smallskip
\begin{minipage}{\linewidth}
\footnotesize
\textbf{Note:} $d=1$ denotes the initial candidate generated during breadth exploration, while $d=2,\ldots,5$ denote successive refinement levels. The percentages are calculated over all correct repairs for each model, and each bug is counted at the depth where its first correct patch is obtained.
\end{minipage}
\endgroup
\end{table}

{As shown in Table~\ref{tab:lpr-refinement-depth}, although most correct patches are obtained directly at $d=1$, depth refinement makes a substantial contribution to the overall repair results. Specifically, 50 of the 236 correct repairs with \ModelName{}$_{\text{GPT-4}}$ (21.19\%) and 65 of the 186 correct repairs with \ModelName{}$_{\text{GPT-4o}}$ (34.95\%) are obtained only after at least one refinement step. Moreover, refinement gains are concentrated at relatively shallow depths: 84.00\% of the refined repairs with \ModelName{}$_{\text{GPT-4}}$ and 78.46\% with \ModelName{}$_{\text{GPT-4o}}$ are completed within the first two refinement steps ($d\le3$). This distribution shows that updated execution-level feedback can convert a meaningful subset of initially unsuccessful candidates into correct patches without requiring deep refinement.}

{Successful refinement, however, does not necessarily exhibit monotonic improvement in test outcomes. Specifically, 56.00\% of the successful \ModelName{}$_{\text{GPT-4}}$ refinement trajectories and 56.92\% of the \ModelName{}$_{\text{GPT-4o}}$ trajectories contain at least one step in which the number of failing tests does not decrease; 8.00\% and 6.15\%, respectively, even temporarily introduce additional test failures before ultimately reaching a correct patch. This suggests that failing-test count alone is not a reliable monotonic indicator of refinement progress. Meanwhile, 95.46\% of all \ModelName{}$_{\text{GPT-4}}$ refinement trajectories and 91.60\% of \ModelName{}$_{\text{GPT-4o}}$ trajectories end either because a subsequent patch fails compilation or because the predefined depth limit is reached. Taken together, these observations indicate that refinement eligibility does not imply that an intermediate patch is necessarily closer to the correct solution; rather, LPR retains executable intermediate states from which updated runtime information can be collected and explores them within a bounded depth budget.}
\FloatBarrier

\subsubsection{{\textit{Runtime Overhead and End-to-End Repair Time}}}
Dynamic APR techniques inherently introduce runtime overhead due to program instrumentation and trace collection. To quantify this cost, we evaluate the runtime overhead of ByteTrace, the underlying instrumentation mechanism used by \ModelName{}, on Defects4J v2.0, covering both single-function and multi-function bugs. {Specifically, for each bug, we measure the execution time of the same test workload with and without ByteTrace, and define the instrumentation overhead as the difference between the two execution times.}

\begin{figure}[pos=htbp]
    \centering
    \includegraphics[width=0.65\linewidth]{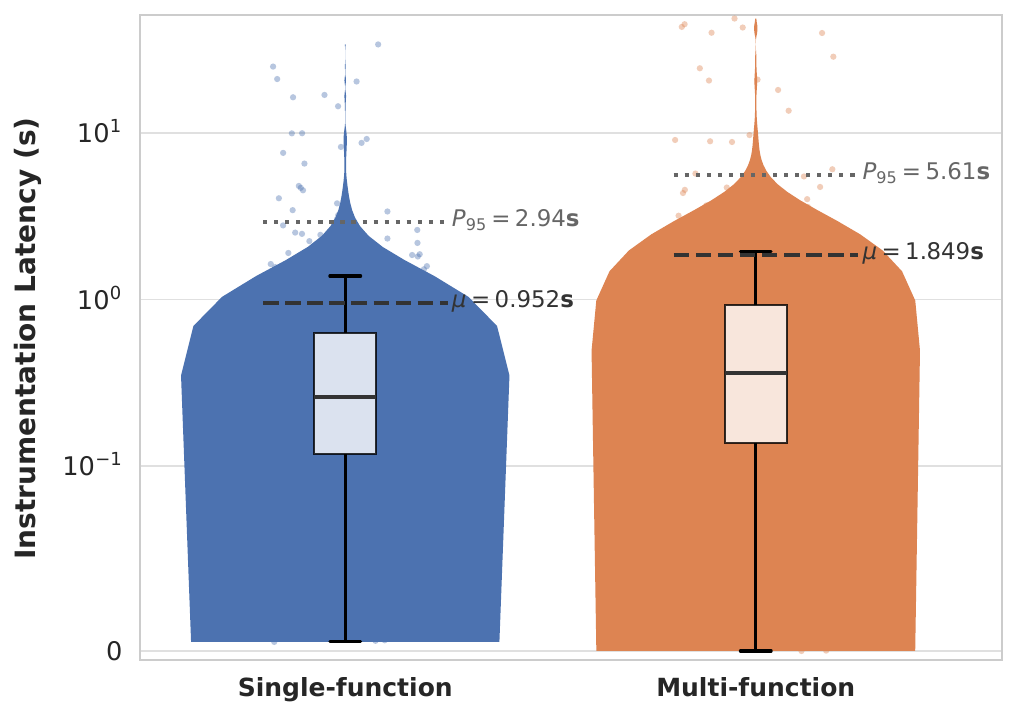}
    \caption{Distribution of {incremental} instrumentation {overhead} for Single-function and Multi-function repairs. Multi-function repairs exhibit higher {additional} runtime overhead ($\mu=1.849\,\mathrm{s}$) compared to Single-function repairs ($\mu=0.952\,\mathrm{s}$). Dashed lines indicate the mean, and dotted lines mark the 95th percentile ($P_{95}$).}
    \label{fig:overhead_comparison}
\end{figure}

\begin{figure}[pos=htbp]
    \centering
    \includegraphics[width=0.6\linewidth]{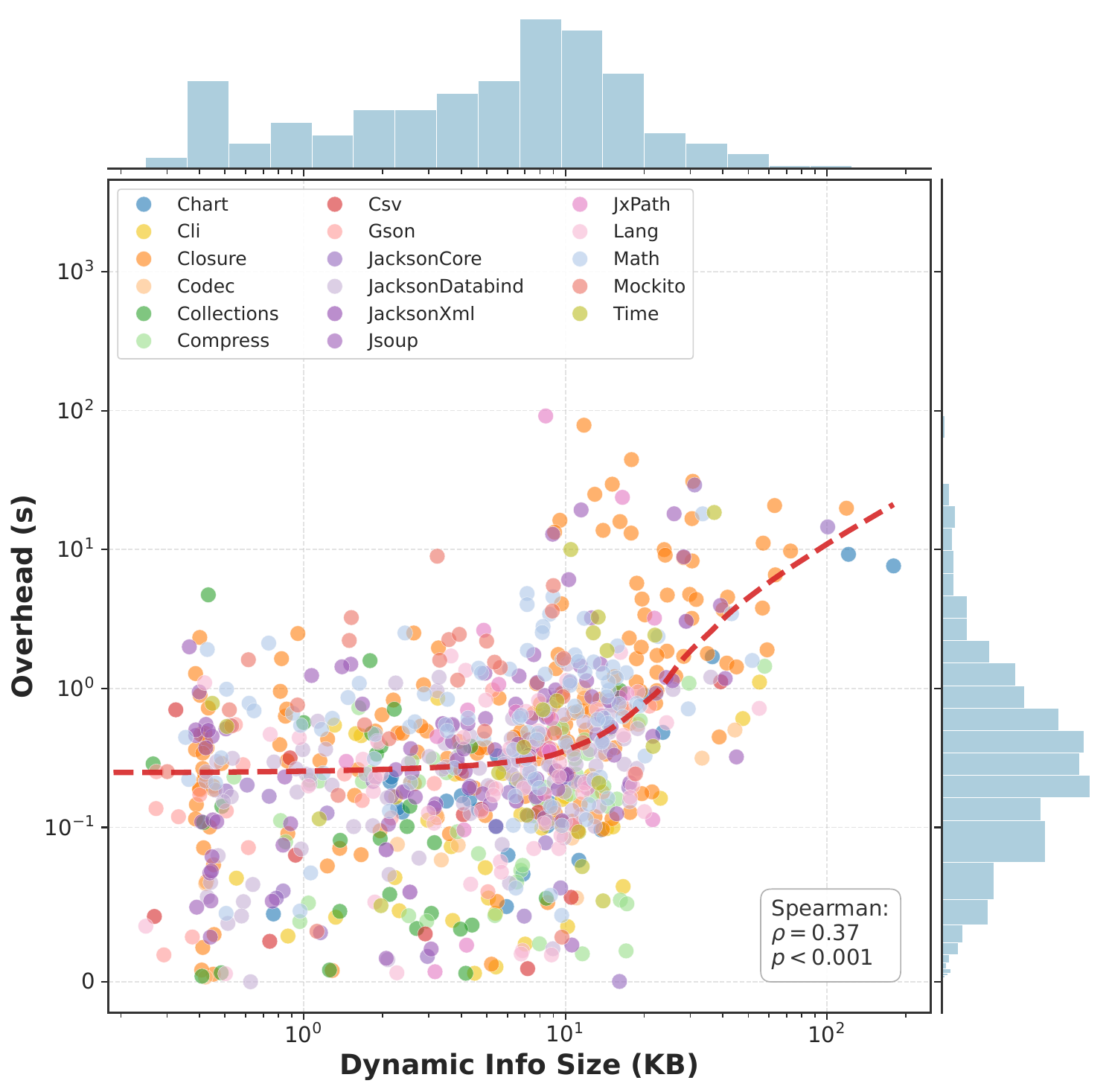}
    \caption{Correlation between dynamic information size and {incremental instrumentation} overhead (Spearman $\rho=0.37, p<0.001$).}
    \label{fig:overhead_trend}
\end{figure}

As shown in Figure~\ref{fig:overhead_comparison}, the average {incremental instrumentation overhead} for single-function bugs is 0.952 seconds, indicating {that ByteTrace introduces limited additional execution time} in most cases. For multi-function bugs, the average {additional overhead} increases to 1.849 seconds, but remains within a practical, second-level range.

We further examine the relationship between instrumentation overhead and the size of the collected execution traces (Figure~\ref{fig:overhead_trend}). {Here, instrumentation overhead refers to the additional execution time introduced by ByteTrace relative to the corresponding uninstrumented test execution.} The observed Spearman correlation coefficient is 0.37 ($p < 0.001$), suggesting a moderate positive correlation. Non-linear regression analysis indicates a growth exponent of approximately 1.5, implying that the overhead follows a low-order polynomial trend, rather than exponential growth.
These results demonstrate that the {incremental instrumentation cost incurred during} dynamic information acquisition scales in a controlled manner. Even when capturing execution contexts exceeding 100 KB, the {additional} instrumentation overhead remains manageable, supporting the feasibility of integrating execution-level feedback in practical repair scenarios.

{We further analyze the end-to-end repair time of \ModelName{} on 388 Defects4J v1.2 bugs, including both single-function and multi-function cases. Whereas Figure~\ref{fig:overhead_comparison} reports the incremental overhead introduced by ByteTrace, Table~\ref{tab:end-to-end-runtime} reports the cumulative Dynamic Information Collection time across repair iterations, including instrumented test execution and trace collection and processing. As shown in Table~\ref{tab:end-to-end-runtime}, the mean, median, and P95 total repair times are 585.900 s, 212.769 s, and 1605.093 s, respectively. Among the major components, compilation and testing account for the largest runtime cost, followed by Dynamic Info. Collection and LLM/API calls, while other framework overhead remains negligible. Notably, the mean repair time of unfixed bugs reaches 808.650 s, substantially higher than the 143.827 s observed for fixed bugs, indicating that unsuccessful cases typically consume a larger portion of the available search budget before termination. Overall, these results show that the end-to-end runtime of \ModelName{} is dominated by validation, dynamic information collection, and iterative search costs rather than framework-level overhead. Since the measured runtime is affected by factors such as LLM/API response latency and machine performance, these results should be interpreted as representative rather than absolute execution times.}

\begin{table}[pos=htbp]
\centering
\caption{{End-to-end runtime breakdown of \ModelName{} on 388 Defects4J v1.2 bugs. Fixed denotes repair attempts that generate at least one plausible patch, whereas Unfixed denotes those that generate no plausible patch.}}
\label{tab:end-to-end-runtime}
\begingroup
\small
\setlength{\tabcolsep}{6pt}
\renewcommand{\arraystretch}{1.15}
\arrayrulecolor{black}
\begin{tabular}{@{}lrrr@{}}
\toprule
Component & Mean (s) & Median (s) & P95 (s) \\
\midrule
Dynamic Info. Collection & 172.246 & 7.442 & 252.789 \\
LLM/API & 121.989 & 63.904 & 397.164 \\
Compile \& Test & 291.315 & 96.007 & 1201.371 \\
Other Overhead & 0.351 & 0.272 & 0.946 \\
\midrule
Total -- All & 585.900 & 212.769 & 1605.093 \\
Total -- Fixed & 143.827 & 39.500 & 856.214 \\
Total -- Unfixed & 808.650 & 458.536 & 2103.640 \\
\bottomrule
\end{tabular}
\arrayrulecolor{black}
\endgroup
\end{table}
\FloatBarrier

\subsubsection{\textit{Token Consumption and Cost Efficiency Analysis}}
\begin{figure}[pos=htbp]
    \centering
    \includegraphics[width=0.6\linewidth]{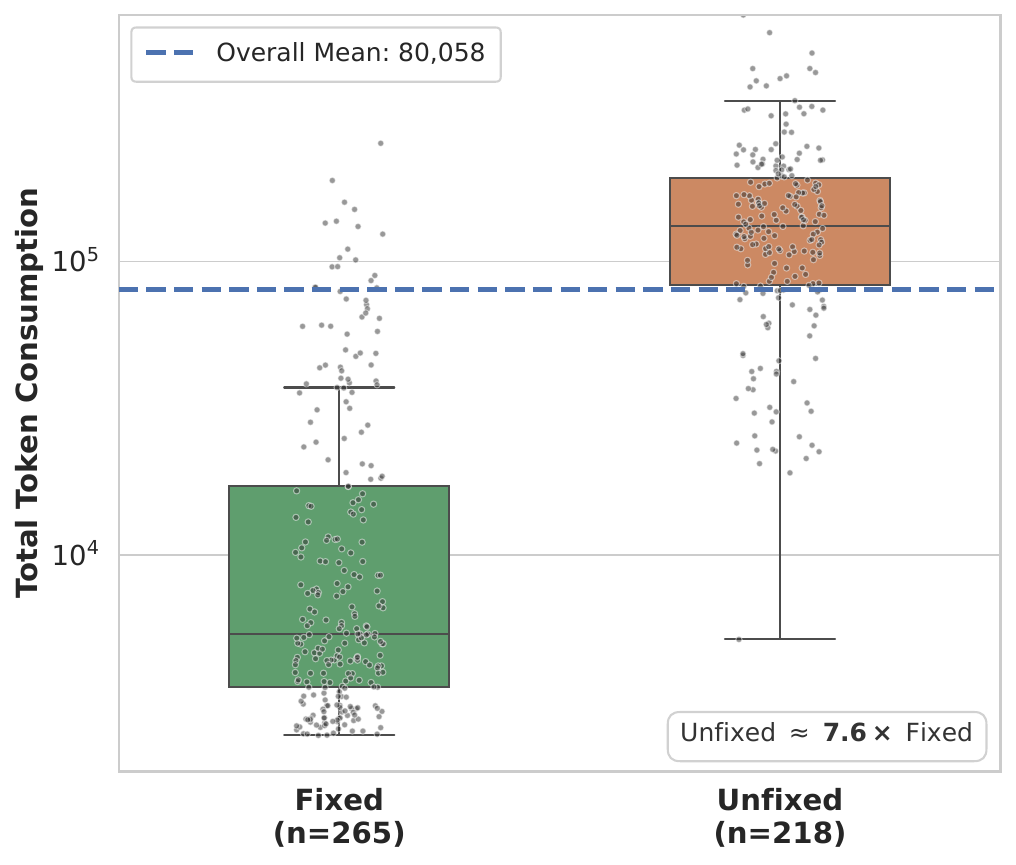}
    \caption{Distribution of {\ModelName{}} token consumption for Fixed and Unfixed repair attempts.
    {Fixed denotes repair attempts that generate at least one plausible patch, whereas Unfixed denotes those that generate no plausible patch.}
    The box plots show that unsuccessful repairs (Unfixed) consume approximately $7.6\times$ as many tokens as Fixed repairs.
    The blue dashed line indicates the overall mean consumption.}
    \label{fig:cost_analysis}
\end{figure}

{In addition to runtime overhead, token consumption plays a critical role in determining the economic feasibility of LLM-based APR techniques. Figure~\ref{fig:cost_analysis} compares the token consumption of Fixed and Unfixed repair attempts on the 483 single-function bugs. On average, \ModelName{} consumes approximately 80,058 tokens per bug. For a coarse comparison with previously reported results, RepairAgent and ChatRepair require approximately 270,000 and 210,000 tokens per bug, respectively, corresponding to reductions of 70.3\% and 61.9\%. Based on current GPT-4o pricing, this translates to an average repair cost of approximately \$0.253 per bug for \ModelName{}, substantially lower than the \$1.45 average cost reported for ReinFix under comparable settings. Figure~\ref{fig:cost_analysis} also shows that Unfixed repair attempts consume approximately $7.6\times$ as many tokens as Fixed repair attempts, highlighting the importance of controlling unproductive search trajectories.}

{To further understand the source of this efficiency, we reproduced RepairAgent and conducted a fine-grained comparison under the same experimental configuration on 388 Defects4J v1.2 bugs, including both single-function and multi-function cases. As shown in Table~\ref{tab:token_fine_grained}, \ModelName{} performs 13.28 LLM calls per bug, compared with 35.82 for RepairAgent, corresponding to a 62.93\% reduction in invocation frequency. Despite incorporating execution-level dynamic information, \ModelName{} also consumes fewer input tokens per call (6,745.48 vs.\ 8,504.18). Overall, its total token consumption decreases from 311,803.50 to 95,633.17 tokens per bug, corresponding to a 69.33\% reduction.}

\begin{table}[pos=htbp]
\centering
\caption{{Fine-grained comparison of LLM invocation and token consumption between DynaFix and RepairAgent on 388 Defects4J v1.2 bugs, including both single-function and multi-function cases.}}
\label{tab:token_fine_grained}
\begingroup

\arrayrulecolor{black}
\resizebox{\columnwidth}{!}{%
\begin{tabular}{@{}lccccccc@{}}
\toprule
Method & \# Bugs & \shortstack{LLM Calls \\ / Bug} & \shortstack{Input Tokens \\ / Call} & \shortstack{Output Tokens \\ / Call} & \shortstack{Input Tokens \\ / Bug} & \shortstack{Output Tokens \\ / Bug} & \shortstack{Total Tokens \\ / Bug} \\
\midrule
DynaFix & 388 & 13.28 & 6,745.48 & 458.11 & 89,551.45 & 6,081.72 & 95,633.17 \\
RepairAgent & 388 & 35.82 & 8,504.18 & 200.65 & 304,616.39 & 7,187.11 & 311,803.50 \\
\bottomrule
\end{tabular}%
}
\arrayrulecolor{black}
\endgroup
\end{table}

{This reduction is mainly driven by a more concentrated repair process. Under the $B=7, D=5$ configuration, \ModelName{} allows at most 35 candidate generations per bug, while the observed average is only 13.28, i.e., 37.94\% of the maximum search budget. In addition, execution traces from previous refinement iterations are not accumulated; each iteration uses the latest execution information from the current candidate patch. Thus, updated execution-level feedback provides targeted runtime guidance without causing the prompt context to continuously grow. The higher average of 95,633.17 tokens per bug in Table~\ref{tab:token_fine_grained}, compared with 80,058 in Figure~\ref{fig:cost_analysis}, results from the broader evaluation scope: Figure~\ref{fig:cost_analysis} covers 483 single-function bugs, whereas Table~\ref{tab:token_fine_grained} includes both single-function and multi-function bugs, which generally require larger execution contexts.}

{Overall, these results show that \ModelName{}'s token efficiency mainly stems from reducing unnecessary LLM invocations and terminating unproductive repair trajectories early, while keeping the iterative execution context bounded.}

\begingroup

\hypersetup{citecolor=black,linkcolor=black}
\subsubsection{{\textit{Comparison with Autonomous Coding Agents}}}

To further examine DynaFix under agentic repair settings, we compare it with RepairAgent and SWE-agent using the controlled configuration described in Section 4.5. RepairAgent serves as the more direct baseline because it was specifically designed for Defects4J repair, whereas SWE-agent was originally developed for repository-level issue resolution on SWE-bench and was adapted to Defects4J while preserving its original interaction workflow as much as possible. Therefore, the SWE-agent results are mainly used to characterize the broader behavior of agentic repair rather than as a fully like-for-like comparison under each system's native benchmark setting. Table~\ref{tab:coding_agent_comparison} summarizes their repair effectiveness, search effort, and overall cost.

\begin{table}[pos=htbp]
\centering

\colorlet{scolor}{black}
\caption{{Comparison of DynaFix with autonomous coding agents on Defects4J v1.2.}}
\label{tab:coding_agent_comparison}
\begingroup

\arrayrulecolor{black}
\renewcommand{\arraystretch}{1.18}
\begin{tabular}{@{}llrrr@{}}
\toprule
Category & Metric & DynaFix & RepairAgent & SWE-agent \\
\midrule
\multirow{2}{*}{Effectiveness} & \#Correct & \textbf{112} & 108 & 90 \\
 & Repair Rate & \textbf{28.87\%} & 27.84\% & 23.20\% \\
\midrule
\multirow{2}{*}{Search Effort} & LLM Calls / Bug & \textbf{13.28} & 35.82 & 25.95 \\
 & Compile \& Test / Bug & 12.88 & 104.85 & \textbf{6.06} \\
\midrule
\multirow{2}{*}{Overall Cost} & Total Tokens / Bug & \textbf{95,633} & 311,803 & 301,575 \\
 & Time / Bug (s) & 585.9 & 3,084.3 & \textbf{200.8} \\
\bottomrule
\end{tabular}
\arrayrulecolor{black}
\endgroup
\end{table}

First, DynaFix achieves the highest repair effectiveness with substantially less LLM-driven exploration. It repairs 112 bugs, compared with 108 for RepairAgent and 90 for SWE-agent, while requiring only 13.28 LLM calls per bug, 62.9\% fewer than RepairAgent and 48.8\% fewer than SWE-agent. DynaFix also consumes only 95,633 tokens per bug, approximately one-third of the consumption of both agents. These results suggest that providing fine-grained runtime states can help reduce repeated LLM interactions for acquiring failure evidence, thereby lowering LLM-mediated search cost.

Second, RepairAgent exhibits a validation-intensive repair process. Although its repair effectiveness is close to DynaFix (108 vs.\ 112 bugs), it requires $2.70\times$ as many LLM calls, $8.14\times$ as many compile-and-test operations, $3.26\times$ as many tokens, and $5.26\times$ as much time. Its high number of compile-and-test operations indicates substantial repeated validation during exploration. In contrast, DynaFix uses execution-level information to guide subsequent repair decisions, which is consistent with its lower observed search effort.

Third, SWE-agent exhibits a different cost profile. It performs only 6.06 compile-and-test operations per bug and completes in 200.8 seconds on average, both lower than DynaFix. However, it requires 25.95 LLM calls and 301,575 tokens per bug, corresponding to $1.95\times$ the LLM calls and $3.15\times$ the tokens of DynaFix, while repairing 22 fewer bugs. This indicates that fewer executions do not necessarily imply lower overall search cost, suggesting that a larger share of the cost may instead arise from LLM-mediated code inspection, reasoning, and modification.

At the same time, DynaFix does not dominate coding agents in every efficiency dimension. Its systematic instrumentation and dynamic-information collection introduce additional execution overhead, as reflected by the lower runtime and fewer compile-and-test operations of SWE-agent. Moreover, general-purpose coding agents can flexibly select what information to inspect and which actions to perform, whereas DynaFix follows a more structured dynamic-analysis workflow. Such flexibility may be advantageous for heterogeneous bugs where the predefined dynamic information does not capture the most relevant evidence. Nevertheless, recent work by \shortcites{xiang2026empowering}\citet{xiang2026empowering} shows that autonomous debugging agents can also benefit from fine-grained dynamic execution information, suggesting that fine-grained dynamic analysis and agentic repair are complementary rather than mutually exclusive.
\par
\endgroup
\FloatBarrier

\begin{tcolorbox}[answerbox]

\noindent \textbf{Answer to RQ3:}
\textit{\ModelName{} achieves a favorable trade-off between repair effectiveness and computational cost. By configuring the LPR strategy with a search breadth of $B=7$ and a search depth of $D=5$, the framework achieves an effective balance between repair capability and computational overhead. {Depth refinement also makes a substantial contribution to the overall repair results: 21.19\% and 34.95\% of the correct repairs with GPT-4 and GPT-4o, respectively, are obtained only after at least one refinement step, with most refined repairs completed within the first two refinement steps. For runtime efficiency, individual instrumentation-and-trace-collection executions have average latencies of 0.952 seconds for single-function bugs and 1.849 seconds for multi-function bugs, while the instrumentation overhead follows a low-order polynomial trend as trace size increases. Across the 388 evaluated Defects4J v1.2 bugs, the mean and median end-to-end repair times are 585.900 and 212.769 seconds, respectively, with the overall runtime dominated by validation and iterative search costs rather than framework-level overhead.} Moreover, guided by execution-level feedback and {by reducing unnecessary LLM invocations and terminating unproductive repair trajectories early}, \ModelName{} reduces token consumption by approximately 60\%--70\% {relative to the reported results of RepairAgent and ChatRepair}, achieving an average repair cost of approximately \$0.253 per bug {for the 483 single-function bugs}. {The controlled comparison with autonomous coding agents further shows that \ModelName{} repairs 112 bugs, compared with 108 for RepairAgent and 90 for SWE-agent, while requiring fewer LLM calls and substantially fewer tokens. At the same time, SWE-agent requires fewer compile-and-test operations and lower runtime, indicating that \ModelName{} does not dominate coding agents in every efficiency dimension. Overall, these results show that \ModelName{} achieves a favorable performance--cost trade-off while maintaining manageable runtime overhead.}}

\end{tcolorbox}

\subsection{RQ4: Impact of Feedback Mechanisms and System Components on Repair Effectiveness}
To better understand the sources of effectiveness in \ModelName{}, this research question investigates how different feedback mechanisms and dynamic information components contribute to repair performance. We first conduct a macro-level analysis to quantify the impact of feedback at increasing levels of sophistication, with particular attention to performance degradation in multi-function scenarios. We then perform a fine-grained ablation study to isolate the contribution of individual execution-level information components.

\subsubsection{\textit{Hierarchical Effectiveness of Feedback Mechanisms}}
To evaluate how progressively richer feedback enhances LLM-based program repair, we conducted experiments on the complete Defects4J v2.0 benchmark, which contains 483 single-function bugs and 347 multi-function bugs. We compared four experimental configurations:
(1) Pure LLM, which relies solely on the buggy code;
(2) LLM + Exception Info, which augments the prompt with standard JUnit exception messages;
(3) LLM + Execution-Level Dynamic Info, which incorporates runtime traces in a single-turn setting; and (4) \ModelName{}, which integrates execution-level dynamic information with an iterative feedback-driven repair process.

\begin{table}[t]
\caption{Effectiveness of DynaFix vs.\ Execution-Level Dynamic Information and Traditional Exception Information in Enhancing LLM-based Program Repair on Defects4J v2.0 (measured by correct fixes).}
\centering
{\small
\begin{tabular}{llcccc}
\toprule
Bug Type & Enhancement Method & \#Bugs & \#Correct & Fix Rate & $\Delta$ over Pure\\
\midrule
\multirow{4}{*}{Single-function}
& DynaFix               & 483 & 186 & 38.5\% & +25.5\% \\
& Execution-Level Info  & 483 & 102 & 21.1\% & +8.1\%  \\
& Exception Info        & 483 & 72  & 14.9\% & +1.9\%  \\
& Pure                  & 483 & 63  & 13.0\% & -       \\
\midrule
\multirow{4}{*}{Multi-function}
& DynaFix               & 347 & 25  & 7.2\%  & +4.9\%  \\
& Execution-Level Info  & 347 & 15  & 4.3\%  & +2.0\%  \\
& Exception Info        & 347 & 13  & 3.7\%  & +1.4\%  \\
& Pure                  & 347 & 8   & 2.3\%  & -       \\
\bottomrule
\end{tabular}%
}
\label{tab:dynafix_vs_exception}
\end{table}

Table~\ref{tab:dynafix_vs_exception} summarizes the repair effectiveness under these configurations.
For single-function bugs, the Pure LLM successfully repaired only 63 cases (13.0\%). Augmenting the prompt with exception messages yielded a modest improvement to 14.9\%, while incorporating execution-level dynamic information further increased the repair rate to 21.1\%. These results indicate that dynamic runtime information is more informative than traditional exception messages, but its benefit remains limited when used in a single-pass manner. In contrast, when dynamic information is coupled with the iterative repair mechanism in \ModelName{}, the number of repaired bugs rises sharply to 186 (38.5\%), representing an absolute improvement of 25.5 percentage points over the Pure LLM. {This substantially outperforms both non-iterative baselines, suggesting that iterative feedback is important for fully exploiting execution-level semantics.}

\citet{xin2024detecting} report that, in multi-function defect scenarios, current APR techniques typically achieve repair rates of only 4\%–7\%, substantially lower than those for single-function bugs. We observe a similar trend in our experiments. In this setting, incorporating execution-level dynamic information in a single-turn manner offers only a marginal advantage over exception information, with overall improvements remaining limited. Nevertheless, once the iterative mechanism is introduced, \ModelName{} repairs 25 bugs (7.2\%), more than tripling the Pure LLM’s performance and substantially outperforming all baseline configurations. This demonstrates that iterative feedback remains beneficial even under challenging multi-function conditions.

\begin{figure}[pos=htbp]
    \centering
    \begin{subfigure}[b]{0.48\linewidth}
        \centering
        \includegraphics[width=\linewidth]{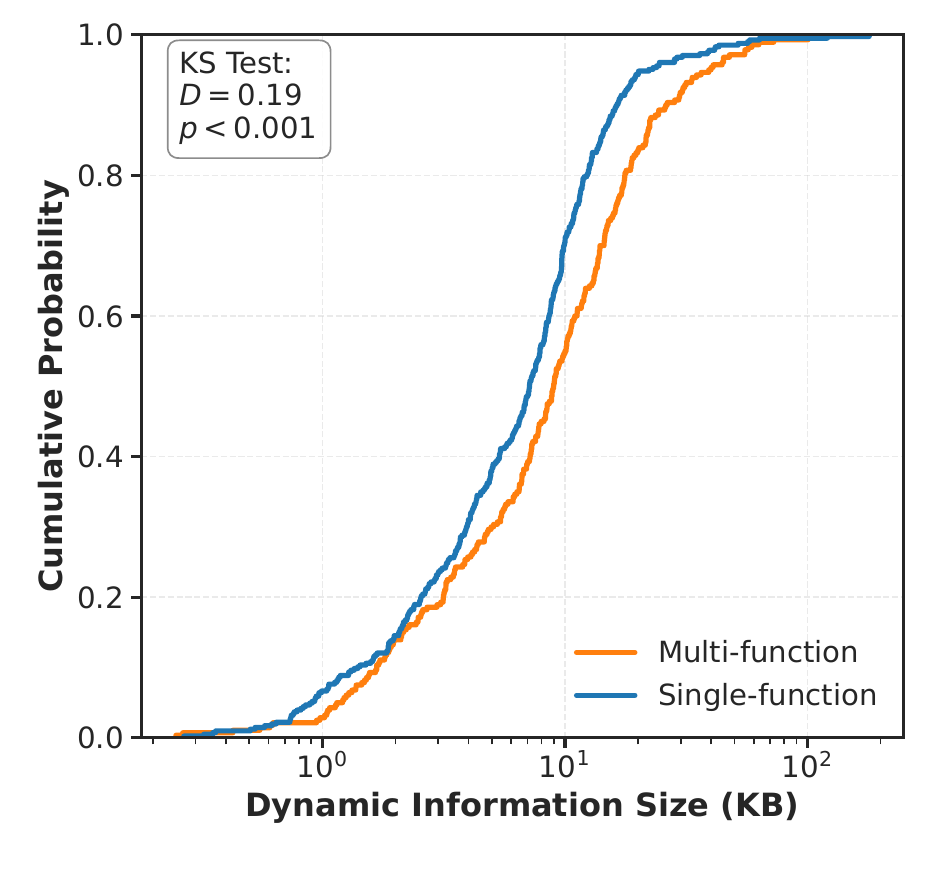}
        \caption{Cumulative Distribution (ECDF)}
        \label{fig:ecdf}
    \end{subfigure}
    \hfill
    \begin{subfigure}[b]{0.48\linewidth}
        \centering
        \includegraphics[width=\linewidth]{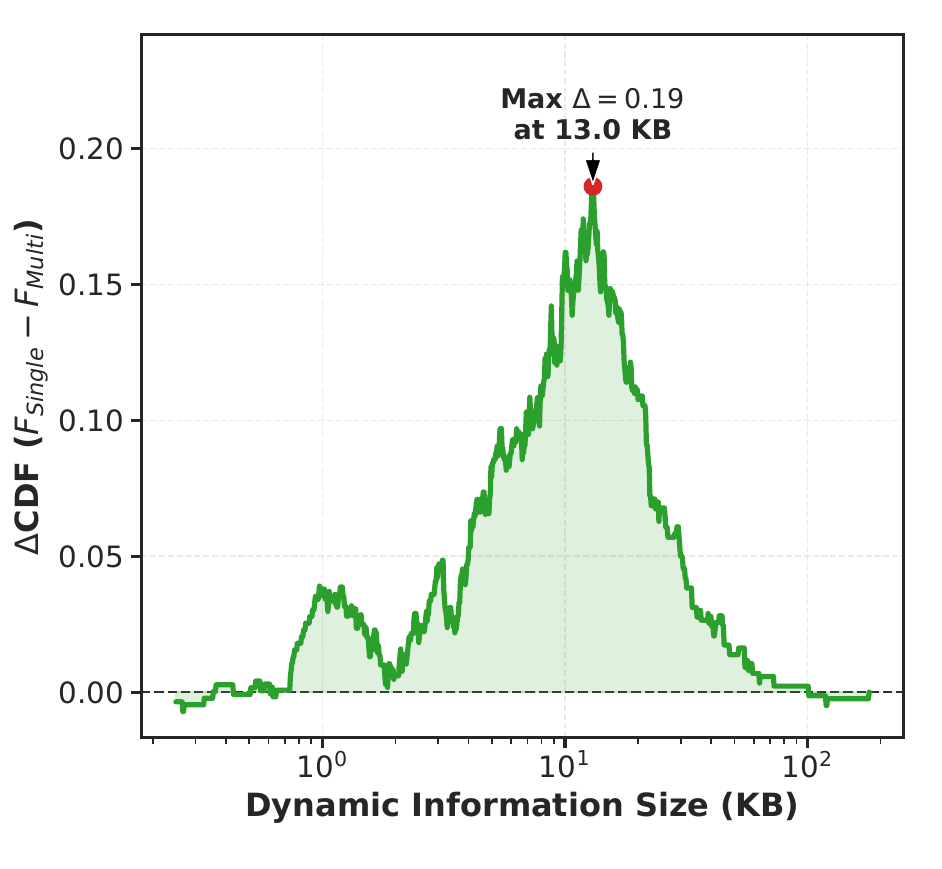}
        \caption{Distributional Difference}
        \label{fig:delta_cdf}
    \end{subfigure}

    \caption{Distributional analysis of dynamic information size. (a) ECDFs reveal that multi-function repairs generally require capturing larger dynamic contexts compared to single-function ones. (b) The distributional difference ($\Delta$CDF) quantifies this gap, showing a maximum divergence of 0.19 at 13.0 KB.}
    \label{fig:dynamic_info_size}
\end{figure}

\subsubsection{\textit{Analysis of the Multi-Function Performance Gap}}
Despite its consistent superiority over all baselines, \ModelName{} exhibits a pronounced performance gap between single-function (38.5\%) and multi-function (7.2\%) scenarios. To investigate this discrepancy, we analyze the distribution of dynamic information sizes, as shown in Figure~\ref{fig:dynamic_info_size}.

The empirical cumulative distribution functions reveal that single-function bugs generally require substantially smaller dynamic contexts than multi-function bugs. The multi-function distribution is shifted toward larger context sizes, indicating heavier information loads. A Kolmogorov--Smirnov test confirms a statistically significant difference between the two distributions ($D=0.19$, $p<0.001$), with the maximum divergence occurring at approximately 13.0 KB.
This distributional shift highlights two compounding challenges. First, larger contexts introduce substantial noise, making it difficult for LLMs to distinguish critical execution dependencies from irrelevant information. Prior studies have shown that as contextual scope increases, models increasingly struggle to attend to mid-context information, which is often where key causal signals reside \citep{liu2023repobench, liu2024lost}. Second, to prevent context window overflow, \ModelName{} applies mandatory truncation, which may remove parts of the execution trace essential for reconstructing complete causal chains. Together, these factors lead to both information overload and partial information loss, substantially constraining repair effectiveness in multi-function scenarios.

% Major revision: separating dynamic information and iterative feedback.
\FloatBarrier
\Needspace{14\baselineskip}
\subsubsection{{\textit{Separating Dynamic Information and Iterative Feedback}}}
\label{sec:rq4-separating-feedback}
\begingroup

To further distinguish the roles of execution-level dynamic information and iterative feedback, we compare four configurations using GPT-4o on the same 255 single-function bugs from Defects4J v1.2. All configurations use the same model, bug set, repair example, and generation budget of at most 35 candidate patches per bug. One-shot Static independently generates candidates using only static code. Iterative Static/Test-feedback iteratively uses previous repair attempts and their test feedback without execution-level dynamic information. One-shot Dynamic additionally uses execution-level information collected from the original buggy program, but does not feed previous candidates or validation outcomes into subsequent generations. Iterative Dynamic (DynaFix) combines execution-level information with iterative feedback and recollects updated execution information from the latest program state during refinement. Here, one-shot refers to the absence of a feedback loop between candidate generations rather than a budget of a single candidate.
\par
\endgroup
\begingroup
\def\RQfourInManuscript{1}
\def\RQfourRevisionText#1{{#1}}
\def\RQfourRevisionStyle{\colorlet{scolor}{black}\arrayrulecolor{black}}
\begin{table}[pos=H]
\RQfourRevisionStyle
\centering
\caption{\RQfourRevisionText{Comparison of four repair configurations on 255 single-function bugs from Defects4J v1.2.}}
\label{tab:rq4-four-way}
\begingroup
% Match Table 7: checkmark for enabled components; en dash for disabled components.
\footnotesize
\setlength{\tabcolsep}{7pt}
\renewcommand{\arraystretch}{1.12}
\resizebox{\linewidth}{!}{%
\begin{tabular}{lccccc}
\toprule
\textbf{Configuration} & \textbf{Static} & \textbf{Dynamic Info} & \textbf{Iterative Feedback} & \textbf{Max Patches} & \textbf{\#Correct} \\
\midrule
One-shot Static & \checkmark & -- & -- & 35 & 70 \\
Iterative Static/Test-feedback & \checkmark & -- & \checkmark & 35 & 71 \\
One-shot Dynamic & \checkmark & \checkmark & -- & 35 & 87 \\
Iterative Dynamic (DynaFix) & \checkmark & \checkmark & \checkmark & 35 & 100 \\
\bottomrule
\end{tabular}%
}
\endgroup
\end{table}
\arrayrulecolor{black}
\endgroup
\FloatBarrier

\begingroup

As shown in Table~\ref{tab:rq4-four-way}, the four configurations correctly repair 70, 71, 87, and 100 bugs, respectively. Compared with One-shot Static, Iterative Static/Test-feedback increases the number of correct repairs from 70 to 71, corresponding to 1 additional bug (+1.4\%). One-shot Dynamic increases the number from 70 to 87, corresponding to 17 additional bugs (+24.3\%). This result shows that execution-level information provides a substantial benefit even without iterative feedback. DynaFix further repairs 100 bugs, outperforming One-shot Dynamic by 13 bugs (+14.9\%) and Iterative Static/Test-feedback by 29 bugs (+40.8\%).

These results support the distinction between using execution-level information once and continuously integrating updated execution information into an iterative repair loop. Fixed execution information already improves repair effectiveness, whereas DynaFix re-executes the modified program as candidate patches evolve, refreshes the runtime information, and uses the resulting execution behavior to guide subsequent refinement. The distinguishing design of DynaFix therefore lies not merely in introducing execution-level dynamic information, but in turning it from a one-time auxiliary context into continuously updated feedback throughout the repair process.
\par
\endgroup

\subsubsection{\textit{Ablation Study of Dynamic Information Components}}
\begin{table}[tbp]
\centering
\caption{Repair results (correct fixes) of \ModelName{} when using different dynamic information components on Defects4J v1.2.}
\label{tab:ablation-sensitivity}
\renewcommand{\arraystretch}{1.2}
{\small
\begin{tabular}{lccccc}
\toprule
\multicolumn{1}{c}{\multirow{2}{*}{\textbf{Variants}}}
& \multicolumn{3}{c}{\textbf{Components}}
& \multicolumn{2}{c}{\textbf{Result}} \\
\cmidrule(lr){2-4} \cmidrule(l){5-6}
& Local Variable
& Control Flow
& Method Call
& \#Correct
& $\Delta$ Impact (\%) \\
\midrule
DynaFix$_{Full}$
& \checkmark & \checkmark & \checkmark
& \multicolumn{1}{c}{100} & -- \\
\midrule
w/o Local Variable
& -- & \checkmark & \checkmark
& 85 ($-$15) & $-$15.0\% \\
w/o Control Flow
& \checkmark & -- & \checkmark
& 86 ($-$14) & $-$14.0\% \\
w/o Method Call
& \checkmark & \checkmark & --
& 91 ($-$9) & $-$9.0\% \\
\midrule
DynaFix$_{Base}$
& -- & -- & --
& 71 ($-$29) & $-$29.0\% \\
\bottomrule
\end{tabular}
}
\end{table}

To quantify the individual contributions of each execution-level information component, we conducted an ablation study on 255 single-function bugs from Defects4J v1.2. Using the optimal search parameters identified in RQ3, we evaluated variants of \ModelName{} by selectively removing specific types of dynamic information:
\begin{itemize}
    \item \textbf{w/o Local Variables:} excludes runtime values of local variables;
    \item \textbf{w/o Control Flow:} excludes execution path and branch information;
    \item \textbf{w/o Method Call:} excludes call stack and invocation relationships;
    \item \textbf{DynaFix$_{\text{Base}}$:} uses only the buggy function code without dynamic information;
    \item \textbf{DynaFix$_{\text{Full}}$:} the complete configuration incorporating all dynamic components.
\end{itemize}

Table~\ref{tab:ablation-sensitivity} reports the corresponding repair results. Overall, all dynamic information components contribute positively to repair performance.
Compared with DynaFix$_{\text{Base}}$, which correctly repairs 71 bugs,
the full configuration correctly repairs 100 bugs, indicating a substantial improvement enabled by execution-level semantics.

Among the individual components, both local variable information and control flow traces play major roles. Removing local variable values results in a 15.0\% reduction in repair effectiveness, while removing control flow information leads to a comparable drop of 14.0\%. These results highlight the importance of concrete runtime states and execution ordering in guiding LLM reasoning. In addition, removing method call information also causes a noticeable performance decrease of 9.0\%, suggesting that invocation structure provides complementary semantic constraints. Overall, no single dynamic component is sufficient on its own; instead, \ModelName{} derives its effectiveness from the synergistic integration of multiple sources of execution-level information.

\begin{tcolorbox}[answerbox]

\noindent \textbf{Answer to RQ4:}\textit{
\ModelName{}'s effectiveness arises from the combined impact of an iterative feedback mechanism and complementary execution-level dynamic information.
{Execution-level dynamic information improves repair effectiveness even without iterative feedback, while continuously updating and integrating such information throughout iterative repair yields further improvements.}
Ablation results further show that local variable states, method invocation relationships, and control flow traces {all contribute positively to repair effectiveness}, collectively forming a robust and effective set of semantic constraints for LLM-based program repair.
}

\end{tcolorbox}

\section{Threats to Validity}

\noindent \textbf{Data Leakage.}
A potential threat to internal validity arises from the possibility that Large Language Models may have been exposed to correct patches or similar bug–fix pairs during pre-training, which could inflate repair performance. To mitigate this risk, we conducted an additional evaluation of \ModelName{} on the RWB, as reported in \textit{Section~5.2.1}. The RWB consists of real-world bugs collected after the documented training cutoff dates of the evaluated LLMs, substantially reducing the likelihood that these instances appeared in the models’ pre-training corpora. As shown in \textit{Section~5.2.1}, \ModelName{} exhibits performance trends on the RWB that are consistent with those observed on Defects4J, suggesting that its effectiveness does not rely on memorization of previously seen bug–patch patterns. While this evaluation cannot completely eliminate the risk of data leakage, it provides evidence that such leakage is unlikely to be the primary driver of the reported results.

\noindent \textcolor{revisionblue}{\textbf{Manual Patch Correctness Assessment.} As described in \textit{Section~3.3}, two authors independently evaluated plausible patches and resolved disagreements through discussion. However, since only the final consensus-level annotations were retained across experimental settings, inter-rater agreement statistics (e.g., Cohen's $\kappa$) could not be reliably computed. This limitation prevents us from quantifying the consistency of individual judgments and may introduce subjectivity into the reported patch correctness assessments.}

\noindent \textbf{Language Specificity.}
Our empirical evaluation focuses exclusively on Java programs, which introduces a potential threat to external validity regarding the generalizability of \ModelName{} to other programming languages. Nevertheless, the core design of \ModelName{}---capturing execution-level dynamic information through program instrumentation and integrating it into an iterative repair loop---is not inherently tied to Java. Similar instrumentation and tracing mechanisms are available for many mainstream programming languages. {Specifically, the high-level repair components of \ModelName{}, including prompt construction, LLM interaction, patch validation, and the LPR strategy, can be largely reused across programming languages. In contrast, the current ByteTrace implementation relies on JVM bytecode instrumentation and therefore needs to be replaced by a language-specific tracing backend. For example, in Python, a ByteTrace-like backend could use \texttt{sys.settrace()} to capture function call/return events and line-level execution traces, while local-variable states can be obtained from runtime frames. These signals can then be converted into the same variable-state, control-flow, and function-call information used by \ModelName{}. Other language-dependent interfaces, such as program execution, testing, and patch application, may also require adaptation.} Despite this, differences in language semantics, runtime environments, and tooling ecosystems may affect the ease of adaptation and practical effectiveness. {Therefore, extending \ModelName{} to another language mainly requires adapting the tracing and execution-related interfaces rather than redesigning the core repair workflow.} We therefore acknowledge that cross-language generalization remains to be empirically validated, and we leave a comprehensive multi-language evaluation to future work.

\noindent \textcolor{revisionblue}{\textbf{Cross-Model-Family Generalizability.} The main effectiveness evaluation in RQ1 primarily relies on GPT-4 and GPT-4o. Although RQ2 provides additional evidence across the GPT and DeepSeek model families, including comparisons on the same Defects4J-Trans benchmark, these experiments use different benchmarks and configurations from RQ1. Therefore, while these results demonstrate DynaFix's effectiveness under the evaluated model families and settings, the extent to which the main findings of RQ1 generalize to other model families remains to be further validated.}

\noindent \textbf{Information Overload in Multi-function Repair.}
As discussed in \textit{Section~5.4.1}, \ModelName{} achieves lower absolute repair rates on multi-function bugs than on single-function bugs. Further analysis in \textit{Section~5.4.2} indicates that this limitation is largely attributable to the extensive execution traces produced by inter-procedural interactions. These traces substantially increase the input context length, which may lead to information overload and reduce the model’s ability to effectively attend to critical execution signals. Although \ModelName{} mitigates this issue through truncation mechanisms, some loss of relevant information is unavoidable. Addressing this challenge remains an open research problem. In future work, we plan to explore hierarchical indexing and summarization-based context compression techniques to reduce contextual noise while preserving salient debugging information, with the goal of improving repair effectiveness for complex multi-function scenarios.

\noindent {\textbf{Representativeness of Code Perturbations.}
Defects4J-Trans provides a controlled setting for evaluating repair robustness under semantics-preserving changes in source representation. However, the transformations in this benchmark may not capture the full diversity and distribution of source-level changes that arise during real-world software maintenance. Therefore, our results demonstrate repair stability under controlled semantics-preserving perturbations, while robustness to naturally occurring software evolution requires further empirical validation.}

\section{Related Work}
APR has been extensively studied from multiple methodological perspectives. Traditional approaches primarily rely on predefined repair templates~\citep{martinez2016astor,long2017automatic,hua2018towards,liu2019avatar,ghanbari2019practical,jiang2019inferring}, heuristic search and rule-based techniques~\citep{le2011genprog,le2016history,long2016automatic,xin2017leveraging,xiong2017precise,wen2018context,jiang2018shaping}, and constraint- or synthesis-based formulations~\citep{xuan2016nopol,le2017s3,long2015staged,mechtaev2016angelix}. More recently, deep learning–based methods have been proposed to learn repair patterns from data~\citep{chen2019sequencer,lutellier2020coconut,li2020dlfix,zhu2021syntax,jiang2021cure,ye2022neural,zhu2023tare}, significantly expanding the design space of APR systems.

With the emergence of large language models (LLMs), a growing body of work has explored LLM-based APR. \emph{ReinFix}~\citep{zhang2025repair} analyzes bug causes by mining internal and external repair ingredients to guide patch generation.
\emph{ThinkRepair}~\citep{yin2024thinkrepair} improves repair performance by enhancing LLM reasoning through chain-of-thought prompting.
\emph{ChatRepair}~\citep{xia2024automated} formulates APR as a dialogue-driven process, iteratively refining patches using test failure feedback and historical repair attempts.
\emph{GIANTRepair}~\citep{li2025giantrepair} constrains the search space by extracting patch skeletons from LLM outputs and instantiating them with contextual information.
\emph{RepairAgent}~\citep{bouzenia2025repairagent} models the LLM as an autonomous agent that plans and executes repair actions via dynamic prompts and a state machine.
\emph{FitRepair}~\citep{xia2023plastic} incorporates the plastic surgery hypothesis through patch knowledge and repair-oriented fine-tuning, while
\emph{SelfAPR}~\citep{ye2022selfapr} and
\emph{Self-Debug}~\citep{chen2023teaching} leverage compiler diagnostics, test feedback, and chain-of-thought reasoning to improve self-supervised repair or generation quality.
Despite their effectiveness, most existing LLM-based APR techniques predominantly rely on static program representations, such as source code, test failures, and error messages, or only utilize coarse-grained runtime signals (e.g., pass/fail outcomes). As a result, the rich semantic information embedded in fine-grained execution behavior remains largely underexploited.

Several recent studies have begun to explore the use of execution-level dynamic information for APR. \emph{TraceFixer}~\citep{bouzenia2023tracefixer} incorporates local variable values and expected execution states during CodeT5 fine-tuning to enhance repair accuracy. \emph{Towards Effectively Leveraging Execution Traces}~\citep{haque2025towards} systematically analyzes the role of execution traces in APR and demonstrates their utility for explaining failing test behaviors. Similarly, \emph{LDB}~\citep{zhong2024debug} partitions programs into basic blocks, tracks intermediate variable values, and employs LLMs for step-by-step verification of program states. {\emph{InspectCoder}~\citep{wang2025inspectcoder} uses debugger breakpoints to selectively inspect specific runtime states at suspicious program locations and leverages these observations to support program repair. \citet{farzandway2025automated} use spectrum-based fault localization (SBFL) to derive suspiciousness scores from test execution coverage and leverage the ranked suspicious code locations to guide LLM-based repair.}

{In contrast to these approaches, \ModelName{} repeatedly updates and exploits execution-level dynamic information throughout iterative repair through an automated instrumentation-and-re-execution loop. When a candidate patch remains refinable after validation, \ModelName{} re-executes the updated program, recollects structured execution information, and feeds the updated runtime evidence into the next repair iteration, so that the feedback remains aligned with the evolving candidate program.}

\section{Conclusion and Future Work}
In this paper, we presented \ModelName{}, an LLM-based automated program repair framework that {continuously updates and exploits execution-level dynamic information within a closed-loop iterative repair process}. By leveraging the ByteTrace tool for runtime data collection and an LPR strategy, \ModelName{} enables LLMs to reason more effectively about program behavior and to generate context-aware patches within a unified debugging-and-repair workflow.
Experimental results on the Defects4J benchmark demonstrate that \ModelName{} produces 236 correct fixes in total, including 23 bugs that were previously unsolved by existing approaches. Further analyses show that the framework exhibits strong robustness under code perturbations and maintains consistent performance on unseen real-world scenarios. At the same time, \ModelName{} achieves substantial cost efficiency, with an average token consumption of only 80,058 per bug, {reducing token consumption by approximately 60\%–70\% compared with representative LLM-based APR approaches.} These results indicate that systematically integrating execution-level dynamic feedback into automated program repair better aligns LLM-based repair with real-world debugging practices and significantly improves both effectiveness and reliability in complex repair tasks.

In future work, we plan to extend \ModelName{} along three main directions. First, we aim to generalize the dynamic trace collection mechanism to additional programming languages beyond Java, such as Python and C++, thereby broadening the applicability of the framework. Second, we will explore enhancements to support multi-file and multi-hunk bug repair, which require more sophisticated context management and dependency analysis across larger code scopes. {Third, we plan to explore trace-aligned differential feedback between consecutive repair attempts, which may reduce redundant runtime information and more explicitly capture behavioral changes introduced by the latest patch.}

\section*{CRediT authorship contribution statement}

\textbf{Zhili Huang:} Writing -- original draft, Conceptualization, Formal analysis, Methodology, Software, Validation, Visualization.
\textbf{Ling Xu:} Writing -- review \& editing, Supervision.
\textbf{Chao Liu:} Writing -- review \& editing.
\textbf{Weifeng Sun:} Writing -- review \& editing.
\textbf{Xu Zhang:} Writing -- review \& editing.
\textbf{Yan Lei:} Writing -- review \& editing.
\textbf{Hongyu Zhang:} Writing -- review \& editing, Supervision.

\section*{Declaration of generative AI and AI-assisted technologies in the manuscript preparation process}
During the preparation of this work the author(s) used OpenAI Codex and ChatGPT to assist with finding related works and polishing the writing. After using these tools/services, the author(s) reviewed and edited the content as needed and take(s) full responsibility for the content of the published article.

\section*{Declaration of competing interest}
The authors declare that they have no known competing financial interests or personal relationships that could have appeared to influence the work reported in this paper.

\section*{Data Availability}
To facilitate reproducibility and future research, we have made our comprehensive replication package publicly available. {This package includes the source code of the DynaFix framework and all experimental datasets and scripts used in this study. The DynaFix artifacts are hosted on GitHub at \url{https://github.com/gujiprogram/DynaFix}, while the ByteTrace instrumentation tool is released independently at \url{https://github.com/gujiprogram/ByteTrace}.}

\section*{Acknowledgements}
This research was partially funded by the Chongqing Technology Innovation and Application Development Project (No. CSTB2023TIAD-STX0021) and the National Natural Science Foundation of China (No. 62272072). The funders had no role in study design, data collection and analysis, decision to publish, or preparation of the manuscript. We gratefully acknowledge the support of all funding agencies.

\printcredits

% Bibliography embedded for Editorial Manager submission.
% This avoids relying on the submission system to run BibTeX.

\end{document}